\documentclass[onecolumn]{aa}

\bibpunct{(}{)}{;}{a}{}{,}

\usepackage{euclid}
\usepackage{graphicx}
\usepackage{natbib}
\usepackage{scalerel}

\usepackage[table]{xcolor}

\usepackage[varg]{txfonts}

\usepackage[pdfencoding=auto,psdextra]{hyperref}
\hypersetup{
colorlinks=true,
linkcolor=blue,
filecolor=magenta,      
urlcolor=blue,
citecolor=blue
}
\makeatletter
\renewcommand*\aa@pageof{, page \thepage{} of \pageref*{LastPage}}
\makeatother

\usepackage[utf8]{inputenc}
\usepackage{subcaption}
\usepackage[switch, modulo]{lineno}

\usepackage{amsmath}	
\usepackage{dsfont}
\usepackage{newtxtext,newtxmath}
\usepackage{tikz}
\usetikzlibrary{angles, quotes}
\usetikzlibrary{shapes.arrows,calc}
\usepackage{physics} 
\usepackage{multirow}
\usepackage{tabularx}
\usepackage{tikz-cd}
\usepackage{sidecap}

\defcitealias{Linke2022b}{L+23}
\defcitealias{planck2020}{Planck Collaboration VI 2020}
\defcitealias{Euclid_2023}{Euclid Collaboration 2023}
\defcitealias{euclid2025}{Euclid Collaboration 2025}

\begin{document}

\title{A roadmap to cosmological parameter analysis with third-order shear statistics}
\subtitle{IV. Analytic cross-covariance between second- and third-order aperture masses}

\newcommand{\orcid}[1]{} 
\author{
Niek Wielders\inst{\ref{inst1},\ref{inst2}}\thanks{E-mail: niek.wielders@gmail.com}\and
Laila Linke\inst{\ref{inst1},\ref{inst3}}\and
Pierre A. Burger\inst{\ref{inst1},\ref{inst4},\ref{inst5}}\and
Sven Heydenreich\inst{\ref{inst1},\ref{inst6}}\and
Lucas Porth\inst{\ref{inst1}}\and
Peter Schneider\inst{\ref{inst1}}}

\institute{University of Bonn, Argelander-Institut f\"ur Astronomie, Auf dem H\"ugel 71, 53121 Bonn, Germany\label{inst1} \and School of Mathematics, Statistics and Physics, Newcastle University, Newcastle upon Tyne, NE1 7RU, UK\label{inst2} \and Universit\"at Innsbruck, Institut f\"ur Astro- und Teilchenphysik, Technikerstr. 25/8, 6020 Innsbruck, Austria\label{inst3} \and Waterloo Centre for Astrophysics, University of Waterloo, Waterloo, Ontario N2L 3G1, Canada\label{inst4} \and Department of Physics and Astronomy, University of Waterloo, Waterloo, Ontario N2L 3G1, Canada\label{inst5} \and Department of Astronomy and Astrophysics, University of California, Santa Cruz, 1156 High Street, Santa Cruz, CA 95064, USA\label{inst6}}

\date{Date Received: 11 March 2025; Date Accepted: 26 August 2025}

\abstract {Weak gravitational lensing is a powerful probe of cosmology, with second-order shear statistics commonly used to constrain parameters such as the matter density $\Omega_\mathrm{m}$ and the clustering amplitude $S_8$. However, degeneracies between parameters persist and can be broken by including higher-order statistics, such as the third-order aperture mass. To jointly analyse second- and third-order statistics, an accurate model of their cross-covariance is essential.} {This work derives and validates a non-tomographic analytical model for the cross-covariance between second- and third-order aperture mass statistics. Analytical models are computationally efficient and enable cosmological parameter inference across a range of models, in contrast to numerical covariances derived from simulations or resampling methods, which are either costly or biased.} {We derived the cross-covariance from real-space estimators of the aperture mass. Substituting the \texttt{Halofit} power spectrum, \texttt{BiHalofit} bispectrum, and a halo-model-based tetraspectrum, the model was validated against numerical covariances from the $N$-body Scinet LIghtCone Simulations (SLICS) using both shear catalogues and convergence maps. We performed a Markov chain Monte Carlo parameter analysis using both analytical and numerical covariances for several filter scale combinations.} {The cross-covariance separates into three terms governed by the power spectrum, bispectrum, and tetraspectrum, with the latter dominating. While the analytical model qualitatively reproduces simulation results, differences arise due to modelling approximations and numerical evaluation issues. The analytical contours are systematically tighter, with a combined figure of merit that is 72\% that of the numerical case, increasing to 80\% when small-scale information is excluded. These differences largely stem from an underprediction of the second-order covariance.} {This work completes the analytical covariance framework for second- and third-order aperture mass statistics, enabling joint parameter inference without the need for large simulation suites. While further refinement is needed to improve quantitative accuracy, the model represents a key step towards simulation-independent cosmic shear analyses.}

\keywords{gravitational lensing: weak – methods: analytical – large-scale structure of Universe - methods: statistical - cosmological parameters}

\titlerunning{Analytical Cross-Covariance}
\authorrunning{N. Wielders et al.}
\maketitle

\section{Introduction}
The Lambda cold dark matter ($\mathrm{\Lambda CDM}$) model is the prevailing framework for understanding the evolution of the Universe, from the post-inflationary epoch to the development of the large-scale structure (LSS) observed today. It is characterised by six key parameters that must be empirically determined. Observational efforts to constrain these parameters fall into two broad categories: those that examine the early Universe through measurements of the cosmic microwave background (CMB), such as those provided by Planck \citepalias{planck2020}, and those that probe the late-time Universe, notably via the LSS.

A longstanding area of interest in recent years has been the apparent tension between CMB-based and LSS-based determinations of the clustering amplitude parameter $S_8=\sigma_8  \sqrt{\Omega_\mathrm{m}/0.3}$, where $\sigma_8$ quantifies the normalisation of the matter power spectrum and $\Omega_\mathrm{m}$ denotes the total matter density. Several LSS surveys have reported lower values of $S_8$ compared to those inferred from the CMB, with tensions reaching $2-3\sigma$ significance \citep{divalentino2021, Joudaki_2020, Asgari_2021}. However, recent results from the KiDS-Legacy analysis \citep{wright_2025} have shown that this tension can be resolved through a more robust statistical treatment of the data, suggesting that the discrepancy does not require new physics, but rather a refined methodological framework.

Nevertheless, the potential of weak gravitational lensing as a tool for probing cosmology remains undiminished. Through the measurement of background galaxy distortions induced by the intervening matter distribution, weak lensing offers a direct and unbiased probe of the total matter content, encompassing both baryonic and dark components, without assumptions about galaxy bias. This makes it uniquely suited for constraining key cosmological parameters such as $\Omega_\mathrm{m}$ and $S_8$ ($\sigma_8$).

In the past decade, weak lensing analyses have primarily employed second-order statistics, such as the two-point shear correlation function, which have substantially improved parameter constraints \citep{Hildebrandt:2017, DES2022, Heymans2021, Dalal2023}. However, second-order measures alone are limited in their ability to fully characterise the matter distribution, particularly at late times. The Gaussian approximation valid in the early Universe breaks down due to non-linear structure formation, which induces non-Gaussian features such as high-density peaks corresponding to galaxy clusters. These complexities are not captured by second-order statistics alone.

Moreover, degeneracies, especially between $\Omega_\mathrm{m}$ and $\sigma_8$, remain a significant challenge when relying solely on two-point statistics (\citealt{Takada_2004, Kilbinger2005}; \citetalias{Euclid_2023}). To break these degeneracies and access the richer information encoded in the LSS, higher-order statistics and non-Gaussian probes have become an increasingly active area of development. These methods hold the potential not only to tighten parameter constraints, but also to offer a more complete statistical description of the cosmic matter distribution.

In the following, we focus on third-order statistics, which provide a measure of the skewness of the matter distribution. Since second- and third-order statistics have different dependences on $\Omega_\mathrm{m}$ and $S_\mathrm{8}$ \citep{Takada_2004, Kilbinger2005, Kayo2013, Pyne2021}, combining them in a joint parameter analysis helps to break degeneracies, as demonstrated by \citet{Heydenreich2023}.
Various higher-order statistics have been introduced to address the limitations of second-order statistics in weak gravitational lensing. Examples include the shear three-point correlation function \citep{Schneider_2003, Schneider_2005}, peak-count statistics \citep{Martinet:2018,Harnois-Deraps:2021}, persistent homology \citep{Heydenreich:2021}, and density split statistics \citep{Gruen:2018,Burger_2022}.
For our analysis, we adopted second- and third-order aperture mass statistics, $\expval{\mathcal{M}_\mathrm{ap}^2}$ and $\expval{\mathcal{M}_\mathrm{ap}^3}$, due to several advantages. Firstly, they are scalars, which effectively compress data and simplify handling \citep{Heydenreich2023}. Secondly, they are unaffected by the mass-sheet degeneracy, a transformation that adds a constant $1-\lambda$ to the dimensionless surface mass density, while linearly scaling it by $\lambda$, leaving lensing observables unchanged \citep{Falco:1985,1995A&A...294..411S}. Thirdly, they naturally decompose E- and B-modes, where B-modes should not be present in gravitational lensing to leading order, providing a valuable check for systematic errors. Lastly, they can be analytically modelled in a comparatively simple manner.

This paper is part of a series that presents a comprehensive approach to cosmological parameter estimation using third-order shear statistics. The series begins with an in-depth analysis of the analytical model for $\expval{\mathcal{M}_\mathrm{ap}^3}$ \citep{Heydenreich2023}. The second paper derives the analytical auto-covariance for $\expval{\mathcal{M}_\mathrm{ap}^3}$ \citep[][\citetalias{Linke2022b} hereafter]{Linke2022b}. The third paper \citet{Porth2024} introduces an efficient method for computing aperture mass statistics by measuring the shear three-point correlation function (3PCF). Finally, the last paper \citep{Burger2024} presents the first numerical cosmological parameter analysis of the fourth data release of the Kilo-Degree Survey (KiDS-1000), incorporating second- and third-order shear statistics, intrinsic alignments, and the effects of baryonic feedback.
The current paper establishes the foundation for an analytical joint analysis of $\expval{\mathcal{M}_\mathrm{ap}^2}$ and $\expval{\mathcal{M}_\mathrm{ap}^3}$ by deriving the final missing component: the analytical cross-covariance. Once derived, we validated the cross-covariance through a joint parameter analysis on simulated data. To achieve this, we used the analytical third-order auto-covariance from \citetalias{Linke2022b} and the analytical second-order auto-covariance computed in \citet{linke2024supersample}. By combining these covariances with the cross-covariance, we were able to construct posterior distributions using Markov chain Monte Carlo (MCMC) sampling.

Using an analytical covariance instead of a numerical one provides several advantages. A numerical covariance can be obtained either from simulations or directly from observational data. In the case of simulations, the number of realisations required must significantly exceed the number of components in the data vector, making the approach computationally expensive. Running these simulations and computing the covariance is particularly demanding, as data vectors are often large. In contrast, an analytical model significantly reduces computational cost while maintaining accuracy.
Alternatively, the covariance matrix can be estimated using jackknife resampling or bootstrapping. These methods divide the data field into multiple patches that are small enough to ensure there are more patches than data vector components, and compute the sample covariance over these patches. However, this approach has limitations. It neglects correlations on scales larger than the patch size and assumes that the patches are mutually independent, which is not strictly true since adjacent patches share boundaries. As a result, this method introduces a bias in the covariance estimate, whereas an analytical covariance remains unaffected.
Moreover, an analytical approach enables a straightforward evaluation of the covariance under different cosmological models. This is crucial for parameter inference and forecasting, as observational data by definition only samples a single cosmology, and generating large suites of simulations for multiple cosmologies would be prohibitively expensive.

The paper is organised as follows. Section~\ref{sec:Theoretical background} provides a concise review of aperture mass statistics. Section~\ref{sec: cross-cov} introduces real-space estimators for aperture mass statistics and derives the analytical cross-covariance from these estimators. Additionally, we present a numerical method for validating the analytical results. Section~\ref{Sect:model validation} compares the analytical and numerical cross-covariances, where the analytical model is constructed by adapting the publicly available third-order modelling code \texttt{threepoint}.\footnote{\url{https://github.com/sheydenreich/threepoint}} Section~\ref{sec: Cosmological Param} outlines the joint cosmological parameter estimation and its validation against numerical contours. Finally, Sect.~\ref{sec: disc} discusses our findings and presents concluding remarks.

\section{Theoretical background}
\label{sec:Theoretical background}
This section briefly overviews the key quantities of weak gravitational lensing and aperture masses. A more detailed discussion can be found in \citet{Schneider:1996}, \citet{Bartelmann:2001}, \citet{Hoekstra:2008} or \citet{Bartelmann:2010}. This paper assumes a flat universe, such that the comoving angular-diameter distance equals the comoving radial distance $f_K(\chi)=\chi$. Furthermore, we use the flat-sky approximation. The matter density contrast is $\delta(\chi\pmb{\vartheta},\chi)=\rho(\chi\pmb{\vartheta},\chi)/\Bar{\rho}(\chi)-1$, where $\rho(\chi\pmb{\vartheta},\chi)$ is the matter density at angular position $\pmb{\vartheta}$ and comoving distance $\chi$ on the backward light cone, and $\Bar{\rho}(\chi)$ is the average density at $\chi$. Defining the Fourier transform of variables with a tilde, the polyspectra of the density contrast $\mathcal{P}_n^\mathrm{(3d)}$ are defined by
\begin{equation}
    \Big\langle \tilde{\delta}(\pmb{\ell}_1/\chi,\chi) \dots \tilde{\delta}(\pmb{\ell}_n/\chi,\chi) \Big\rangle_\mathrm{c} = (2\pi)^3 \, \delta_\mathrm{D}(\pmb{\ell}_1/\chi+\dots+\pmb{\ell}_n/\chi) \, \mathcal{P}_n^\mathrm{(3d)}(\pmb{\ell}_1/\chi,\dots,\pmb{\ell}_n/\chi;\chi) \;,
\end{equation}
where $\langle \dots \rangle_\mathrm{c}$ are cumulants or connected correlation functions, and $\delta_\mathrm{D}$ is a Dirac delta function.

\subsection{The convergence}
The convergence, or the dimensionless surface mass density, can be calculated from $\delta$ by a weighted line-of-sight integration
\begin{equation}
    \kappa(\pmb{\vartheta}) = \frac{3H_0^2\Omega_\mathrm{m}}{2c^2}\int_0^\infty \dd{\chi}\; q(\chi)\, \chi\, \frac{\delta(\chi\pmb{\vartheta}, \chi)}{a(\chi)}\;, \quad\text{with the weight function}\quad  q(\chi) = \int_\chi^\infty \dd{\chi'}\; p(\chi')\, \frac{\chi'-\chi}{\chi'}\;,
    \label{eq: convergence}
\end{equation}
the Hubble constant $H_0$, the probability distribution of source galaxies in comoving distance $p(\chi)\, \dd\chi$, the speed of light $c$, and the scale factor $a(\chi)$, normalised to unity today. The $n\mathrm{th}$-order polyspectra $\mathcal{P}_n$ of the convergence are defined by
\begin{equation}
    \Big\langle\kappa(\pmb{\vartheta}_1)\dots\kappa(\pmb{\vartheta}_n)\Big\rangle_\mathrm{c}=\int \frac{\dd^2\ell_1}{(2\pi)^2}\dots\frac{\dd^2\ell_n}{(2\pi)^2} \; \mathcal{P}_n(\pmb{\ell}_1,\dots,\pmb{\ell}_n) \, (2\pi)^2 \, \delta_\mathrm{D}\left(\pmb{\ell}_1+\dots+\pmb{\ell}_n\right) \, \mathrm{e}^{-\mathrm{i}\,(\pmb{\ell}_1\cdot\pmb{\vartheta}_1+\dots+\pmb{\ell}_n\cdot\pmb{\vartheta}_n)}\;.
    \label{eq: polyspectra}
\end{equation}
Here and in the following, integrals without explicit boundaries are integrals over $\mathbb{R}^2$. Particularly, the power spectrum $P$, the bispectrum $B$, and the tetraspectrum $P_5$ are the convergence polyspectra important for our present discussion and are defined as
\begin{align}
    P(\ell) = \mathcal{P}_2(\pmb{\ell}, -\pmb{\ell})\;, \quad    B(\pmb{\ell}_1, \pmb{\ell}_2) = \mathcal{P}_3(\pmb{\ell}_1, \pmb{\ell}_2, -\pmb{\ell}_1-\pmb{\ell}_2)\;, \quad \textrm{and } \quad
    P_5(\pmb{\ell}_1, \pmb{\ell}_2, \pmb{\ell}_3, \pmb{\ell}_4) = \mathcal{P}_5(\pmb{\ell}_1, \pmb{\ell}_2, \pmb{\ell}_3, \pmb{\ell}_4, -\pmb{\ell}_1-\pmb{\ell}_2-\pmb{\ell}_3-\pmb{\ell}_4)\;,
\end{align}
where we made use of the homogeneity and isotropy of the Universe. Under the Limber approximation, they are related to the polyspectra of $\delta$ ($\mathcal{P}_n^\mathrm{(3d)}$) by a weighted integration along the line of sight \citep{Limber:1954,Kaiser1997,Kayo2013}:
\begin{equation}
\label{eq: limber}
    \mathcal{P}_n(\pmb{\ell}_1,\dots,\pmb{\ell}_n)=\left(\frac{3H_0^2\Omega_\mathrm{m}}{2c^2}\right)^n\,\int_0^\infty \dd{\chi}\; \frac{q^n(\chi)}{\chi^{n-2}\, a^n(\chi)}\, \mathcal{P}_n^\mathrm{(3d)}(\pmb{\ell}_1/\chi,\dots,\pmb{\ell}_n/\chi; \chi)\;.
\end{equation}

The convergence defined in Eq.~\eqref{eq: convergence} and its $n\mathrm{th}$-order connected correlation function (Eq.~\ref{eq: polyspectra}) will be used to calculate the aperture mass in the following subsection, from which our statistics are derived. If shape noise from the intrinsic ellipticity of galaxies, $\epsilon_\mathrm{i}$, is present, it must be accounted for by adding $\sigma_\mathrm{\epsilon_\mathrm{i}}^2 / (2N)$ to the power spectrum $P(\ell)$, where $\sigma_\mathrm{\epsilon_\mathrm{i}}$ is the two-component intrinsic ellipticity dispersion and $N$ is the galaxy number density. A proof of this correction can be found in Appendix B of \citetalias{Linke2022b}. Since shape noise is Gaussian, it does not affect the bispectrum or tetraspectrum.

\subsection{Definition of the aperture mass}
\label{sec: aperture mass}

The aperture mass $\mathcal{M}_\mathrm{ap}$ can be calculated by either a weighted two-dimensional integration over the convergence or the tangential shear $\gamma_\mathrm{t}$. It was first introduced in \cite{Schneider:1996}, and can be defined by either
\begin{equation}
    \mathcal{M}_\mathrm{ap}(\pmb{\phi},\theta):=\int \dd^2\vartheta \; U_{\theta}\left(\lvert\pmb{\phi}-\pmb{\vartheta}\rvert\right) \, \kappa(\pmb{\vartheta}) \quad \textrm{or} \quad \mathcal{M}_\mathrm{ap}(\pmb{\phi},\theta):=\int \dd^2\vartheta \; Q_{\theta}\left(\lvert\pmb{\phi}-\pmb{\vartheta}\rvert\right) \, \gamma_\mathrm{t}(\pmb{\vartheta})\;,
    \label{eq: map def}
\end{equation}
where  $\pmb{\phi}$ is the aperture centre coordinate, $\theta$ the filter radius, and $U_\theta$ and $Q_{\theta}$ are radially symmetric filter functions. $U_\theta$ is compensated, obeying the relation
\begin{equation}
    \int_0^\infty \dd \vartheta \; \vartheta \, U_\theta(\vartheta) = 0\;, \quad \text{and is connected to $Q_\theta$ through}\quad  Q_\theta(\vartheta)=\frac{2}{\vartheta^2}\int_0^\vartheta \dd\vartheta' \; \vartheta' \, U_\theta(\vartheta')-U_\theta(\vartheta)\;.
    \label{eq: filter U}
\end{equation}
For our purpose, we use the exponential filter function proposed in \cite{Crittenden2002},
\begin{equation}
    U_\theta(\vartheta)=\frac{1}{2\pi\theta^2} \, \left(1-\frac{\vartheta^2}{2\theta^2}\right) \, \mathrm{e}^{-\vartheta^2/(2\theta^2)}\;, \quad \text{such that} \quad Q_\theta(\vartheta)=\frac{\vartheta^2}{4\pi\theta^4}\,\mathrm{e}^{-\vartheta^2/(2\theta^2)}\;.
\end{equation}
The Fourier transform of $U_\theta$ is
\begin{equation}
    \tilde{U}_\theta (\ell)=\frac{\ell^2\theta^2}{2} \, \mathrm{e}^{-\ell^2\theta^2/2}=:\tilde{u}(\ell\theta)\;. 
    \label{eq: uHat}
\end{equation}

The second- and third-order aperture mass statistics are essential for our present discussion and are defined, respectively, by
\begin{equation}
    \label{eq: map formal def}
    \expval{\mathcal{M}_\mathrm{ap}^2} (\theta_1,\theta_2):=\Big\langle \mathcal{M}_\mathrm{ap} (\pmb{\phi},\theta_1) \, \mathcal{M}_\mathrm{ap}(\pmb{\phi},\theta_2)\Big\rangle \quad \textrm{and} \quad \expval{\mathcal{M}_\mathrm{ap}^3} (\theta_1,\theta_2,\theta_3):=\Big\langle  \mathcal{M}_\mathrm{ap}(\pmb{\phi},\theta_1) \, \mathcal{M}_\mathrm{ap}(\pmb{\phi},\theta_2) \, \mathcal{M}_\mathrm{ap}(\pmb{\phi},\theta_3)\Big\rangle\;,
\end{equation}
where $\langle\dots\rangle$ denotes the spatial average over $\pmb{\phi}$. We note that since $\langle\mathcal{M}_\mathrm{ap}\rangle=0$, as the expectation value of the density contrast $\delta$ is by definition zero, the second- and third-order cumulants of the aperture mass are equal to the spatial averages: $\langle\mathcal{M}_\mathrm{ap}^3\rangle_\mathrm{c}=\langle\mathcal{M}_\mathrm{ap}^3\rangle$ and $\langle\mathcal{M}_\mathrm{ap}^2\rangle_\mathrm{c}=\langle\mathcal{M}_\mathrm{ap}^2\rangle$.
Inserting the definition of the aperture mass calculated from the convergence (Eq.~\ref{eq: map def}) into the Eqs.~\eqref{eq: map formal def}, using that the spatial averages are equal to the cumulants and substituting the cumulants of the convergence with the polyspectra as defined in Eq.~\eqref{eq: polyspectra}, one finds
\begin{align}
    \expval{\mathcal{M}_\mathrm{ap}^2}(\theta_1,\theta_2)&=\int \dd^2\vartheta_1 \; U_{\theta_1}\left(\lvert\pmb{\phi}-\pmb{\vartheta}_1\rvert\right) \int \dd^2\vartheta_2 \; U_{\theta_2}\left(\lvert\pmb{\phi}-\pmb{\vartheta}_2\rvert\right) \int\frac{\dd^2\ell_1}{(2\pi)^2} \int\frac{\dd^2\ell_2}{(2\pi)^2} \; \mathcal{P}_2(\pmb{\ell}_1,\pmb{\ell}_2) \, (2\pi)^2 \, \delta_\mathrm{D}(\pmb{\ell}_1+\pmb{\ell}_2) \, \mathrm{e}^{-\mathrm{i}\,(\pmb{\ell}_1\cdot\pmb{\vartheta}_1+\pmb{\ell}_2\cdot\pmb{\vartheta}_2)} \nonumber \\ &= \int\frac{\dd^2\ell_1}{(2\pi)^2} \int\frac{\dd^2\ell_2}{(2\pi)^2} \; \tilde{u}(\ell_1 \theta_1) \, \tilde{u}(\ell_2 \theta_2) \, \mathcal{P}_2(\pmb{\ell}_1,\pmb{\ell}_2) \, (2\pi)^2 \, \delta_\mathrm{D}(\pmb{\ell}_1+\pmb{\ell}_2) \, \mathrm{e}^{-\mathrm{i}\, (\pmb{\ell}_1+\pmb{\ell}_2)\cdot \pmb{\phi}}
    \label{eq: Map2_fourier} \\
    &=\int\frac{\dd^2\ell}{(2\pi)^2}\; \tilde{u}(\ell \theta_1)\,\tilde{u}(\ell \theta_2) \, P(\ell)\,, \nonumber
\end{align}
for the second-order aperture mass statistics, where going from the first to the second line, we performed the Fourier transformation on the $U_\theta$ filters. In the last step, we integrated the Dirac delta function. It is important to note that taking equal aperture radii for the second-order aperture mass statistics (i.e., $\theta_1=\theta_2$) captures all the information in this order. With the chosen $\tilde{u}$ filter (Eq.~\ref{eq: uHat}), the dispersion at two distinct aperture radii can be expressed as a dispersion at an average aperture radius \citep{Schneider_2005}
\begin{equation}
    \left\langle\mathcal{M}_\mathrm{ap}(\theta_1)\,\mathcal{M}_\mathrm{ap}(\theta_2)\right\rangle=\int\frac{\dd^2\ell}{(2\pi)^2}\;P(\ell) \,\frac{l^4\theta_1^2\theta_2^2}{4} \, \mathrm{e}^{-l^2\left(\theta_1^2+\theta_2^2\right)/2}=\frac{4\theta_1^2\theta_2^2}{\left(\theta_1^2+\theta_2^2\right)^2}\left\langle\mathcal{M}_\mathrm{ap}^2\left(\sqrt{\frac{\theta_1^2+\theta_2^2}{2}}\right)\right\rangle\,.
    \label{eq: equal radii}
\end{equation}
Therefore, using equal aperture radii in the dispersion preserves all relevant information, and it is sufficient to evaluate $\expval{\mathcal{M}_\mathrm{ap}^2}(\theta):=\expval{\mathcal{M}_\mathrm{ap}^2}(\theta,\theta)$ with a single aperture radius as input. Equivalently, for the third-order
\begin{align}
    \expval{\mathcal{M}_\mathrm{ap}^3}(\theta_1,\theta_2,\theta_3)&=\int \frac{\dd^2\ell_1}{(2\pi)^2}\int \frac{\dd^2\ell_2}{(2\pi)^2}\int \frac{\dd^2\ell_3}{(2\pi)^2} \; \tilde{u}(\ell_1\theta_1) \, \tilde{u}(\ell_2\theta_2) \, \tilde{u}(\ell_3\theta_3)  \, \mathcal{P}_3(\pmb{\ell}_1,\pmb{\ell}_2,\pmb{\ell}_3)  \, (2\pi)^2 \, \delta_\mathrm{D}(\pmb{\ell}_1+\pmb{\ell}_2+\pmb{\ell}_3) \, \mathrm{e}^{-\mathrm{i}\,(\pmb{\ell}_1+\pmb{\ell}_2+\pmb{\ell}_3)\cdot\pmb{\phi}} \nonumber\\
    &=\int \frac{\dd^2\ell_1}{(2\pi)^2}\int \frac{\dd^2\ell_2}{(2\pi)^2} \; \tilde{u}(\ell_1\theta_1) \, \tilde{u}(\ell_2\theta_2) \, \tilde{u}(\abs{\pmb{\ell}_1+\pmb{\ell}_2}\theta_3) \, B(\pmb{\ell}_1,\pmb{\ell}_2)\;.
    \label{eq: Map3_fourier}
\end{align}
The simplification to equal aperture radii used above is not applicable to the third-order statistics. This can be observed from Eq.~\eqref{eq: uHat}, which sharply peaks around $\ell \theta = \sqrt{2}$. For $\expval{\mathcal{M}_\mathrm{ap}^3}$, if one were to take $\theta_1=\theta_2=\theta_3$ in Eq.~\eqref{eq: Map3_fourier}, only equilateral configurations of the bispectrum with $\ell_1 \sim \ell_2 \sim \abs{\pmb{\ell}_1+\pmb{\ell}_2}$ would be probed \citep{Schneider_2005}. However, the non-equilateral aperture radii configurations depend on different bispectrum configurations and provide additional information in this order. Therefore, distinct aperture radii should be allowed for third-order statistics.

\section{Derivation of the analytical cross-covariance}
\label{sec: cross-cov}
\begin{figure}
\begin{minipage}[c]{0.30\linewidth}
\centering
\begin{tikzpicture}
    \draw (0,0) rectangle (5,5);
    \draw (5,0) node[anchor=south east]{$A'$};
    \draw [fill=pink](1,1) rectangle (4,4);
    \draw (4,1) node[anchor=south east]{$A$};
    
    \filldraw [black] (1.5, 2) circle (2pt);
    \draw (1.5, 2) node[anchor=west]{$\pmb{\phi}$};
    \draw (1.5, 2) circle (1);
    \draw (1.5, 2) -- (0.5,2);
    \draw (0.76, 2.05) node[anchor=north]{$4\,\theta$};

    \filldraw [black] (2.5, 4.5) circle (2pt);
    \draw (2.5, 4.5) node[anchor=west]{$\pmb{\phi}'$};
    \draw (2.5, 4.5) circle (1);
    \draw (2.5, 4.5) -- (2.5, 3.5);
    \draw (2.4, 3.75) node[anchor=west]{$4\,\theta$};
    \end{tikzpicture}
\end{minipage}
\begin{minipage}[c]{0.69\linewidth}
\caption{Aperture centre placement in a data field (from \citetalias{Linke2022b}). Given a data field of area $A'$ (not necessarily a square field), apertures are only placed within a smaller field $A$, which lies well within $A'$. The boundary of $A$ is chosen such that 99.9\% of the effective support of $Q_\theta$ lies within $A$ (which corresponds to $\sim 4\,\theta$) if the aperture centre lies on this boundary. To illustrate this, two aperture centres are depicted: $\pmb{\phi}$ is within $A$ and $\pmb{\phi}'$ is outside of $A$. It can be seen that the circle around $\pmb{\phi}'$, illustrating the effective support of $Q_\theta$, extends beyond $A'$, thus introducing bias into the calculated aperture mass at this point. To exclude this effect, we only average the aperture centre over the red region $A$.}
\label{fig: aperture mass estimation}
\end{minipage}
\end{figure}
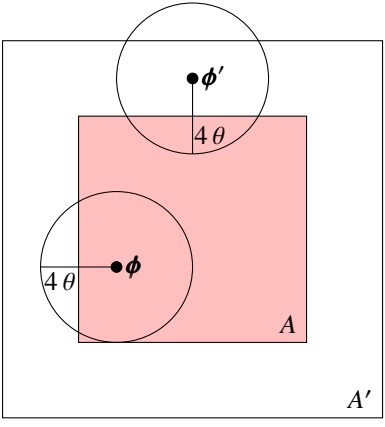

In a real survey or simulation, convolving with the weight function $U_\theta$ over an infinite area is not feasible. Therefore, following \citetalias{Linke2022b}, we define a real-space estimator of the aperture mass as
\begin{equation}
    \hat{\mathcal{M}}_\mathrm{ap}(\pmb{\phi},\theta):=\int_{A'} \dd^2\vartheta \; U_{\theta}\left(\lvert\pmb{\phi}-\pmb{\vartheta}\rvert\right) \, \kappa(\pmb{\vartheta})\;,
    \label{eq: map estimator}
\end{equation}
where $A'$ represents the finite survey area. By spatially averaging products of this real-space estimator over the aperture centre $\pmb{\phi}$, we estimate the second- and third-order aperture mass statistics. However, since we are working with a finite survey area, this spatial average is itself an approximation. Moreover, the averaging region must be smaller than the full survey area, as the aperture mass involves integrating over a region surrounding the aperture centre. Placing aperture centres too close to the survey boundary would introduce biases due to boundary effects. To mitigate these effects, we estimate the second- and third-order aperture mass statistics by averaging over an area $A$,  which is entirely contained within the survey region $A'$ as illustrated in Fig.~\ref{fig: aperture mass estimation}. The area $A$ is chosen such that $99.9\%$ of the effective support of $Q_\theta$ is within $A'$ when the aperture centre $\pmb{\phi}$ is placed on the boundary of $A$ and the largest aperture radius $\theta_\mathrm{max}$ is used. This condition is satisfied when the boundary of $A$ is positioned $4\,\theta_\mathrm{max}$ inside the boundary of $A'$ \citep{Heydenreich2023}. The field boundary of $A$ is determined using $Q_\theta$ rather than $U_\theta$ because $U_\theta$ is a compensated filter function, meaning its support is not well-defined in the same way.
Given this set-up, the integrals over the finite survey area $A'$ can be approximated as integrals over the entire real space $\mathbb{R}^2$, since the convolution with $U_\theta$ contributes negligibly outside of $A'$. The real-space statistics of the aperture mass are thus given by
\begin{equation}
    \hat{\mathcal{M}}_{\mathrm{ap}}^{2}(\theta) = \frac{1}{A} \int_A \dd^2\phi \; \left[ \prod_{i=1}^2 \int \dd^2\vartheta_i \; U_{\theta}\left(\lvert \pmb{\phi}-\pmb{\vartheta}_i\rvert\right) \, \kappa(\pmb{\vartheta}_i)\right] \;\;\text{and}\;\; \hat{\mathcal{M}}_{\mathrm{ap}}^{3}(\theta_1,\theta_2,\theta_3) = \frac{1}{A} \int_A \dd^2\phi \; \left[ \prod_{i=1}^3 \int \dd^2\vartheta_i \; U_{\theta_i}\left(\lvert \pmb{\phi}-\pmb{\vartheta}_i\rvert\right) \, \kappa(\pmb{\vartheta}_i) \right] \;.
    \label{eq: map3 estimator}
\end{equation}    
The cross-covariance of the real-space estimators is then defined as
\begin{equation}
    C_{\hat{\mathcal{M}}_\mathrm{ap}^{3,2}} (\theta_1,\theta_2,\theta_3;\theta_4)=\Big\langle \hat{\mathcal{M}}_{\mathrm{ap}}^{3}\hat{\mathcal{M}}_{\mathrm{ap}}^{2} \Big\rangle(\theta_1,\theta_2,\theta_3;\theta_4)  - \Big\langle \hat{\mathcal{M}}_{\mathrm{ap}}^{3} \Big\rangle (\theta_1,\theta_2,\theta_3)\,\Big\langle \hat{\mathcal{M}}_{\mathrm{ap}}^{2} \Big\rangle(\theta_4)\;.
    \label{eq: cov real-space estimator}
\end{equation}
Here and in the following, when a comma `,' is used, the function is invariant under permutations of its arguments, whereas when a semi-colon `;' is used, they are not invariant. The first term in Eq.~\eqref{eq: cov real-space estimator} is given as
\begin{align}
    \Big\langle\hat{\mathcal{M}}_{\mathrm{ap}}^{3}\hat{\mathcal{M}}_{\mathrm{ap}}^{2}\Big\rangle(\theta_1,\theta_2,\theta_3;\theta_4) =&\; \frac{1}{A^2} \int_A \dd^2\phi_1 \int_A \dd^2\phi_2 \; \left[ \prod_{i=1}^3 \int \dd^2\vartheta_i \; U_{\theta_i}\left(\lvert \pmb{\phi}_1-\pmb{\vartheta}_i\rvert\right) \right] \Bigg[ \prod_{j=4}^5 \int \dd^2\vartheta_j \; U_{\theta_4}\left(\lvert \pmb{\phi}_2-\pmb{\vartheta}_j\rvert\right)\Bigg] \nonumber  \\ &\times \Big\langle\kappa(\pmb{\vartheta}_1)\kappa(\pmb{\vartheta}_2) \kappa(\pmb{\vartheta}_3) \kappa(\pmb{\vartheta}_4) \kappa(\pmb{\vartheta}_5)\Big\rangle \;. 
    \label{eq: cov ensemble}
\end{align}
Here, the fifth-order correlation function can be separated into cumulants,
\begin{align}
    \Big\langle\hat{\mathcal{M}}_{\mathrm{ap}}^{3}\hat{\mathcal{M}}_{\mathrm{ap}}^{2}\Big\rangle(\theta_1,\theta_2,\theta_3;\theta_4) =&\; \frac{1}{A^2} \int_A \dd^2\phi_1 \int_A \dd^2\phi_2 \; \left[ \prod_{i=1}^3 \int \dd^2\vartheta_i \; U_{\theta_i}\left(\lvert \pmb{\phi}_1-\pmb{\vartheta}_i\rvert\right)\right] \Bigg[ \prod_{j=4}^5 \int \dd^2\vartheta_j \; U_{\theta_4}\left(\lvert \pmb{\phi}_2-\pmb{\vartheta}_j\rvert\right)\Bigg] \nonumber  \\ &\times \Bigg\{ \Big\langle\kappa(\pmb{\vartheta}_1) \, \kappa(\pmb{\vartheta}_2)  \, \kappa(\pmb{\vartheta}_3)\Big\rangle_\mathrm{c}  \, \Big\langle\kappa(\pmb{\vartheta}_4)  \, \kappa(\pmb{\vartheta}_5)\Big\rangle_\mathrm{c} \nonumber \\ &+ \Big\langle\kappa(\pmb{\vartheta}_1) \, \kappa(\pmb{\vartheta}_4)  \, \kappa(\pmb{\vartheta}_5)\Big\rangle_\mathrm{c}  \, \Big\langle\kappa(\pmb{\vartheta}_2)  \, \kappa(\pmb{\vartheta}_3)\Big\rangle_\mathrm{c} + \, 2\,\mathrm{Perm.\,of}\,(\theta_1,\theta_2,\theta_3) \label{eq: cov cumulants} \\ &+ 2\Big[\Big\langle\kappa(\pmb{\vartheta}_1) \, \kappa(\pmb{\vartheta}_2)  \, \kappa(\pmb{\vartheta}_4)\Big\rangle_\mathrm{c}  \, \Big\langle\kappa(\pmb{\vartheta}_3)  \, \kappa(\pmb{\vartheta}_5)\Big\rangle_\mathrm{c} + \, 2\,\mathrm{Perm.\,of}\,(\theta_1,\theta_2,\theta_3) \Big] \nonumber \\ &+ \Big\langle\kappa(\pmb{\vartheta}_1)  \, \kappa(\pmb{\vartheta}_2)  \, \kappa(\pmb{\vartheta}_3)  \, \kappa(\pmb{\vartheta}_4)  \, \kappa(\pmb{\vartheta}_5)\Big\rangle_\mathrm{c}\Bigg\} \nonumber
    \\ =:&\; T_\mathrm{PB,1}(\theta_1,\theta_2,\theta_3;\theta_4)+T_\mathrm{PB,2}(\theta_1,\theta_2,\theta_3;\theta_4)+T_\mathrm{PB,3}(\theta_1,\theta_2,\theta_3;\theta_4) + T_\mathrm{P5}(\theta_1,\theta_2,\theta_3;\theta_4)\; , \nonumber
\end{align}
where we defined the $T$-expressions such that $T_\mathrm{PB,1}$ corresponds to the term in the second line, $T_\mathrm{PB,2}$ to the third, $T_\mathrm{PB,3}$ to the fourth, and $T_\mathrm{P5}$ to the fifth. The permutations of $T_\mathrm{PB,2}$ and $T_\mathrm{PB,3}$ can be found in Table~\ref{tab: permutations}.
\begin{table*}
        \caption{Aperture scale radii permutations for $T_\mathrm{PB,2}$ and $T_\mathrm{PB,3}$.}
        \centering
\begin{tabular}{c c}
\hline\hline
$T_\mathrm{PB,2}$ permutations & 
$T_\mathrm{PB,3}$ permutations  \\ \hline
$\theta_1\theta_4\theta_5|\theta_2\theta_3$ & $\theta_1\theta_2\theta_4|\theta_3\theta_5$               \\
$\theta_2\theta_4\theta_5|\theta_1\theta_3$               & $\theta_1\theta_3\theta_4|\theta_2\theta_5$          \\
$\theta_3\theta_4\theta_5|\theta_1\theta_2$               &  $\theta_2\theta_3\theta_4|\theta_1\theta_5$ \\ \hline
\end{tabular}
\tablefoot{The first permutation of each term corresponds to the explicit terms in Eq.~\eqref{eq: cov cumulants}. Interchanging the aperture radii left (right) of the vertical bars is an invariant property.}
        \label{tab: permutations}
\end{table*}
The third term ($T_\mathrm{PB,3}$) includes a factor of two because swapping $\kappa(\pmb{\vartheta}_4)$ with $\kappa(\pmb{\vartheta}_5)$ results in equivalent terms, as both are connected to $\theta_4$. Remember $\langle\kappa\rangle=0$ because the expectation value of the density contrast $\delta$ is zero; thus, the second- and third-order cumulants of the convergence are equal to the respective moments: $\langle\kappa^3\rangle_\mathrm{c}=\langle\kappa^3\rangle$ and $\langle\kappa^2\rangle_\mathrm{c}=\langle\kappa^2\rangle$ (but $\langle\kappa^5\rangle_\mathrm{c}\neq \langle\kappa^5\rangle$). As a result, $T_\mathrm{PB,1}$ cancels the second term on the r.h.s. of Eq.~\eqref{eq: cov real-space estimator}, observing it can be written as the expectation value of $\hat{\mathcal{M}}_\mathrm{ap}^2$ times the expectation value of $\hat{\mathcal{M}}_\mathrm{ap}^3$, such that
\begin{equation}
    C_{\hat{\mathcal{M}}_\mathrm{ap}^{3,2}}(\theta_1,\theta_2,\theta_3;\theta_4)=T_\mathrm{PB,2}(\theta_1,\theta_2,\theta_3;\theta_4) +T_\mathrm{PB,3}(\theta_1,\theta_2,\theta_3;\theta_4) +T_\mathrm{P5}(\theta_1,\theta_2,\theta_3;\theta_4)\;.
    \label{eq: cov estimator}
\end{equation}
The second term $T_\mathrm{PB,2}$ can be rewritten by substituting the definition for the convergence polyspectra (Eq.~\ref{eq: polyspectra}) into Eq.~\eqref{eq: cov cumulants}, cancelling $\pmb{\ell}_3$ and $\pmb{\ell}_5$ by integrating over the Dirac delta functions and performing the Fourier transformations of the $U_\theta$ functions
\begin{align}
    T_\mathrm{PB,2}(\theta_1,\theta_2,\theta_3;\theta_4)=&\;\frac{1}{A^2} \int_A \dd^2\phi_1 \int_A \dd^2\phi_2 \; \left[\prod_{i=1}^3 \int \dd^2\vartheta_i \; U_{\theta_i}\left(\lvert \pmb{\phi}_1-\pmb{\vartheta}_i\rvert\right)\right] \Bigg[\prod_{j=4}^5 \int \dd^2\vartheta_j \; U_{\theta_4}\left(\lvert \pmb{\phi}_2-\pmb{\vartheta}_j\rvert\right)\Bigg]\prod_{k=1}^5\int \frac{\dd^2\ell_k}{(2\pi)^{6}} \nonumber \\ &\times \delta_\mathrm{D}(\pmb{\ell}_2+\pmb{\ell}_3)\, \delta_\mathrm{D}(\pmb{\ell}_1+\pmb{\ell}_4+\pmb{\ell}_5)\, \mathcal{P}_3(\pmb{\ell}_1,\pmb{\ell}_4, \pmb{\ell}_5)  \, \mathcal{P}_2(\pmb{\ell}_2,\pmb{\ell}_3)\, \mathrm{e}^{-\mathrm{i}\,(\pmb{\ell}_1\cdot \pmb{\vartheta}_1+\pmb{\ell}_2\cdot \pmb{\vartheta}_2+\pmb{\ell}_3\cdot \pmb{\vartheta}_3+\pmb{\ell}_4\cdot \pmb{\vartheta}_4+\pmb{\ell}_5\cdot \pmb{\vartheta}_5)} \nonumber \\
     &+ \, 2\,\mathrm{Perm.\,of}\,(\theta_1,\theta_2,\theta_3)\nonumber \\
    =&\; \frac{1}{A^2} \int_A \dd^2\phi_1 \int_A \dd^2\phi_2 \; \left[\prod_{i=1}^3 \int \dd^2\vartheta_i \; U_{\theta_i}\left(\lvert \pmb{\phi}_1-\pmb{\vartheta}_i\rvert\right)\right] \Bigg[ \prod_{j=4}^5 \int \dd^2\vartheta_j \; U_{\theta_4}\left(\lvert \pmb{\phi}_2-\pmb{\vartheta}_j\rvert\right)\Bigg]\,\int \frac{\dd^2\ell_1}{(2\pi)^2}\int \frac{\dd^2\ell_2}{(2\pi)^2}\label{eq: calc}\\ &\times \int \frac{\dd^2\ell_4}{(2\pi)^2}\,B(\pmb{\ell}_1,\pmb{\ell}_4)\, P(\ell_2)\, \mathrm{e}^{-\mathrm{i}\,(\pmb{\ell}_1\cdot (\pmb{\vartheta}_1-\pmb{\vartheta}_5)+\pmb{\ell}_2\cdot (\pmb{\vartheta}_2- \pmb{\vartheta}_3)+\pmb{\ell}_4\cdot (\pmb{\vartheta}_4- \pmb{\vartheta}_5))}+ \, 2\,\mathrm{Perm.\,of}\,(\theta_1,\theta_2,\theta_3) \nonumber \\
    =&\;\frac{1}{A^2} \int_A \dd^2\phi_1 \int_A \dd^2\phi_2\; \int \frac{\dd^2\ell_1}{(2\pi)^2}\int \frac{\dd^2\ell_2}{(2\pi)^2} \int \frac{\dd^2\ell_4}{(2\pi)^2}  \; B(\pmb{\ell}_1,\pmb{\ell}_4)\, P(\ell_2) \nonumber \\ &\times \tilde{u}(\ell_1 \theta_1) \, \tilde{u}(\ell_2 \theta_2)   \, \tilde{u}(\ell_2 \theta_3) \, \tilde{u}(\ell_4 \theta_4) \, \tilde{u}(\lvert\pmb{\ell}_1+\pmb{\ell}_4\rvert \,  \theta_4)\, \mathrm{e}^{-\mathrm{i}\,\pmb{\ell}_1\cdot (\pmb{\phi}_1-\pmb{\phi}_2)}+ \, 2\,\mathrm{Perm.\,of}\,(\theta_1,\theta_2,\theta_3)\;. \nonumber
\end{align}
Collecting all terms containing information on the survey area and geometry in the geometry factor $G_A(\pmb{\ell})$ defined as \citepalias{Linke2022b}
\begin{equation}
    G_A(\pmb{\ell})=\frac{1}{A^2}\int_A \dd^2\phi_1 \int_A \dd^2\phi_2 \; \mathrm{e}^{-\mathrm{i}\,\pmb{\ell}\cdot (\pmb{\phi}_1-\pmb{\phi}_2)}\;,
    \label{eq: geometry}
\end{equation}
we find the final result
\begin{align}
    T_\mathrm{PB,2}(\theta_1,\theta_2,\theta_3;\theta_4)=&\;\int \frac{\dd^2\ell_1}{(2\pi)^2}\int \frac{\dd^2\ell_2}{(2\pi)^2} \int \frac{\dd^2\ell_4}{(2\pi)^2}  \; B(\pmb{\ell}_1,\pmb{\ell}_4)  \, P(\ell_2)  \, \tilde{u}(\ell_1 \theta_1) \, \tilde{u}(\ell_2 \theta_2)   \, \tilde{u}(\ell_2 \theta_3) \, \tilde{u}(\ell_4 \theta_4) \, \tilde{u}(\lvert\pmb{\ell}_1+\pmb{\ell}_4\rvert \,  \theta_4) \nonumber\\ &\times G_A(\pmb{\ell}_1) + 2\, \mathrm{Perm.\,of}\,(\theta_1,\theta_2,\theta_3)\;.
    \label{eq: T2, (3)}
\end{align}
By equivalent calculations, one finds
\begin{align}
    T_\mathrm{PB,3}(\theta_1,\theta_2,\theta_3;\theta_4)=&\;2 \int \frac{\dd^2\ell_1}{(2\pi)^2}\int \frac{\dd^2\ell_2}{(2\pi)^2} \int \frac{\dd^2\ell_3}{(2\pi)^2}  \; B(\pmb{\ell}_2,\pmb{\ell}_3)  \, P(\ell_1) \, \tilde{u}(\ell_1 \theta_1) \, \tilde{u}(\ell_2 \theta_2)  \, \tilde{u}(\ell_3 \theta_3) \, \tilde{u}(\lvert \pmb{\ell}_2+\pmb{\ell}_3\rvert \,  \theta_4) \, \tilde{u}(\ell_1 \theta_4) \nonumber\\ &\times G_A(\pmb{\ell}_1+\pmb{\ell}_2+\pmb{\ell}_3) + 2\, \mathrm{Perm.\,of}\,(\theta_1,\theta_2,\theta_3)\;,
    \label{eq: T3}
\end{align}
and
\begin{align}
    T_\mathrm{P5}(\theta_1,\theta_2,\theta_3;\theta_4)=&\;\int \frac{\dd^2\ell_1}{(2\pi)^2}\int \frac{\dd^2\ell_2}{(2\pi)^2} \int \frac{\dd^2\ell_3}{(2\pi)^2} \int \frac{\dd^2\ell_4}{(2\pi)^2}  \; P_5(\pmb{\ell}_1,\pmb{\ell}_2,\pmb{\ell}_3,\pmb{\ell}_4)  \, \tilde{u}(\ell_1 \theta_1) \, \tilde{u}(\ell_2 \theta_2)  \, \tilde{u}(\ell_3 \theta_3) \, \tilde{u}(\ell_4 \theta_4) \nonumber\\ &\times \tilde{u}(\lvert \pmb{\ell}_1+\pmb{\ell}_2+\pmb{\ell}_3+\pmb{\ell}_4\rvert \, \theta_4)  \, G_A(\pmb{\ell}_1+\pmb{\ell}_2+\pmb{\ell}_3)\;.
    \label{eq: T4}
\end{align}
Equations (\ref{eq: T2, (3)}--\ref{eq: T4}) can subsequently be substituted into Eq.~\eqref{eq: cov estimator} to obtain the analytical covariance of the real-space estimator. Next, we investigate how these terms scale with the survey field size.

\subsection{Large survey area approximation}
This subsection illustrates how the cross-covariance terms behave as the field size increases, by taking the limit $A\to \infty$ in the geometry factor (Eq.~\ref{eq: geometry}). In this limit, one finds \citepalias{Linke2022b}
\begin{equation}
    \lim_{A\to\infty}G_A(\pmb{\ell})=\frac{(2\pi)^2}{A} \, \delta_\mathrm{D}(\pmb{\ell})\;.
    \label{eq: large field approx}
\end{equation}
Substituting and eliminating $\pmb{\ell}_1$ by integrating over the Dirac delta function, the large-field approximation of $T_\mathrm{PB,2}$ becomes
\begin{align}
    T_\mathrm{PB,2}^{\infty}(\theta_1,\theta_2,\theta_3;\theta_4)=&\;\frac{1}{A}\int \frac{\dd^2\ell_2}{(2\pi)^2} \int \frac{\dd^2\ell_4}{(2\pi)^2} \; B(\pmb{0},\pmb{\ell}_4)  \, P(\ell_2) \, \tilde{u}(0) \, \tilde{u}(\ell_2 \theta_2)   \, \tilde{u}(\ell_2 \theta_3) \, \tilde{u}(\ell_4 \theta_4) \, \tilde{u}(\ell_4 \, \theta_4) \nonumber \\ &+ 2\, \mathrm{Permutations\,of}\,(\theta_1,\theta_2,\theta_3) = 0 \;,
    \label{eq: T2, inf}
\end{align}
because $\tilde{u}(0)=0$, as can be seen from Eq.~\eqref{eq: uHat}. Thus, $T_\mathrm{PB,2}$ arises solely from the limited size of the survey field and will therefore be referred to as a finite-field term.

For $T_\mathrm{PB,3}$ and $T_\mathrm{P5}$, as given in Eqs.~\eqref{eq: T3} and \eqref{eq: T4} respectively, the foregoing limiting procedure leads to
\begin{align}
    T_\mathrm{PB,3}^{\infty}(\theta_1,\theta_2,\theta_3;\theta_4)=&\;\frac{2}{A}\int \frac{\dd^2\ell_1}{(2\pi)^2}\int \frac{\dd^2\ell_2}{(2\pi)^2}  \; B(\pmb{\ell}_2,-\pmb{\ell}_1-\pmb{\ell}_2)  \, P(\ell_1) \, \tilde{u}(\ell_1 \theta_1) \, \tilde{u}(\ell_2 \theta_2)  \, \tilde{u}(\lvert \pmb{\ell}_1+\pmb{\ell}_2\rvert \, \theta_3) \, \tilde{u}(\ell_1 \theta_4) \, \tilde{u}(\ell_1 \theta_4) \nonumber\\ &+ 2\, \mathrm{Permutations\,of}\,(\theta_1,\theta_2,\theta_3) \;,
    \label{eq: T3, inf}
\end{align}
and
\begin{equation}
    T_\mathrm{P5}^{\infty}(\theta_1,\theta_2,\theta_3;\theta_4)=\frac{1}{A}\int \frac{\dd^2\ell_1}{(2\pi)^2}\int \frac{\dd^2\ell_2}{(2\pi)^2} \int \frac{\dd^2\ell_4}{(2\pi)^2}  \; P_5(\pmb{\ell}_1,\pmb{\ell}_2,-\pmb{\ell}_1-\pmb{\ell}_2,\pmb{\ell}_4)  \, \tilde{u}(\ell_1 \theta_1) \, \tilde{u}(\ell_2 \theta_2) \, \tilde{u}(\lvert \pmb{\ell}_1+\pmb{\ell}_2\rvert \, \theta_3) \, \tilde{u}(\ell_4 \theta_4) \, \tilde{u}(\ell_4 \theta_4)\;,
    \label{eq: T4, inf}
\end{equation}
such that the cross-covariance in the large-field approximation simplifies to
\begin{equation}
    C_{\hat{\mathcal{M}}_\mathrm{ap}^{3,2}}^{\infty}(\theta_1,\theta_2,\theta_3;\theta_4)= T_\mathrm{PB,3}^{\infty}(\theta_1,\theta_2,\theta_3;\theta_4)+ T_\mathrm{P5}^{\infty}(\theta_1,\theta_2,\theta_3;\theta_4)\;.
\end{equation}
While $T_\mathrm{PB,3}$ and $T_\mathrm{P5}$ scale approximately as $1/A$, $T_\mathrm{PB,2}$ exhibits a steeper decline with increasing field size.

Since the geometry term (Eq.~\ref{eq: geometry}) is a rapidly oscillating function, its integration presents numerical challenges. Because our computational resources are largely sufficient for this task (albeit with some limitations; see Sect.~\ref{sec: disc large}), we retain the full expressions. However, in scenarios with limited computing power, or when computational efficiency is prioritised, large-field approximations may serve as useful alternatives.

A concluding remark on the derived cross-covariance and its large-field approximation is that to compute their values, we estimate $P_5$ in $T_\mathrm{P5}$ using a halo decomposition (Appendix~\ref{app: tetra}). $T_\mathrm{P5}$ is then approximated by keeping the two dominant terms, the one-halo term and one of the two-halo terms:
\begin{equation}
    T_\mathrm{P5}(\theta_1,\theta_2,\theta_3;\theta_4)\approx T_\mathrm{P5,1h}(\theta_1,\theta_2,\theta_3;\theta_4)+T_\mathrm{P5,2h}(\theta_1,\theta_2,\theta_3;\theta_4)\;.
\end{equation}
In the large-field approximation $T_\mathrm{P5,2h}^\infty=0$, which means this is a finite field term, and thus we find $T_\mathrm{P5}^\infty\approx T_\mathrm{P5,1h}^\infty$.

\subsection{Aperture mass correlation functions}
\label{sec: corrf}
This section demonstrates that the covariance terms (Eqs.~\ref{eq: T2, (3)}--\ref{eq: T4}) can alternatively be constructed with aperture mass correlation functions. These correlation functions can be numerically estimated from $N$-body simulations and therefore enable us to verify the individual components of our analytical cross-covariance.

The aperture mass correlation functions are defined as
\begin{multline}
    \mathcal{M}_\mathrm{ap}^{n,m}(\theta_1,\dots,\theta_{m};\theta_{m+1},\dots,\theta_n;\pmb{\eta}):=\Big\langle \mathcal{M}_{\mathrm{ap}}(\pmb{\phi},\theta_1) \dots \mathcal{M}_{\mathrm{ap}}(\pmb{\phi},\theta_m)\,\mathcal{M}_{\mathrm{ap}}(\pmb{\phi}+\pmb{\eta},\theta_{m+1})\dots\mathcal{M}_{\mathrm{ap}}(\pmb{\phi}+\pmb{\eta},\theta_n)\Big\rangle \\=\left[ \prod_{i=1}^m\int\dd^2\vartheta_i \;U_{\theta_1}(\vartheta_1)\dots U_{\theta_m}(\vartheta_m)\right] \Bigg[ \prod_{j=m+1}^n \int\dd^2\vartheta_j \; U_{\theta_{m+1}}\left(\abs{\pmb{\vartheta}_{m+1}+\pmb{\eta}}\right)\dots U_{\theta_{n}}\left(\abs{\pmb{\vartheta}_{n}+\pmb{\eta}}\right)\Bigg]\,\Big\langle\kappa(\pmb{\vartheta}_1)\dots\kappa(\pmb{\vartheta}_n)\Big\rangle\;,
    \label{eq: corrf}
\end{multline}
where, without loss of generality, we set $\pmb{\phi}=\pmb{0}$ after the equality sign by shifting the origin. Substituting the correlation functions into Eq.~\eqref{eq: cov cumulants}, $T_\mathrm{PB,2}$ is rewritten as
\begin{equation}
    T_\mathrm{PB,2}(\theta_1,\theta_2,\theta_3;\theta_4)=\frac{1}{A^2}\mathcal{M}_\mathrm{ap}^{2,0}(\theta_2,\theta_3;\pmb{0}) \int_A \dd^2\phi_1 \int_A \dd^2\phi_2 \; \mathcal{M}_\mathrm{ap}^{3,1}(\theta_1;\theta_4,\theta_4;\pmb{\phi}_1-\pmb{\phi}_2) + 2\, \mathrm{Permutations\,of}\,(\theta_1,\theta_2,\theta_3)\;.
\end{equation}
By replacing the $\pmb{\phi}_2$ integral by an $\pmb{\eta}=\pmb{\phi}_1-\pmb{\phi}_2$ integral and integrating over $\dd[2]\phi_1$, we obtain
$A\,E_A(\pmb{\eta})$, where $E_A(\pmb{\eta})$ represents the probability that for a point $\pmb{\phi}_1$ within $A$, the point $\pmb{\phi}_1+\pmb{\eta}$ also lies within $A$. For a square survey area, $E_A(\pmb{\eta})$ is given in Appendix A.2 of \citet{Heydenreich2020}. By expanding these algebraic steps, we find
\begin{equation}
    T_\mathrm{PB,2}(\theta_1,\theta_2,\theta_3;\theta_4)=\frac{1}{A} \mathcal{M}_\mathrm{ap}^{2,0}(\theta_2,\theta_3;\pmb{0})\int \dd^2\eta \; E_A(\pmb{\eta})  \, \mathcal{M}_\mathrm{ap}^{3,1}(\theta_1;\theta_4,\theta_4;\pmb{\eta}) + 2\, \mathrm{Permutations\,of}\,(\theta_1,\theta_2,\theta_3)\;.
    \label{eq: T_2 final_1}
\end{equation}

After generating the aperture mass correlation function maps from simulations, we azimuthally average them over the polar angle of $\pmb{\eta}$ due to the statistical isotropy of the Universe. Defining the aperture mass correlations as a function of distance $\eta$ by $\bar{\mathcal{M}}_\mathrm{ap}^{n,m}$ (Eq.~\ref{eq: corrf} cannot be used in this context, as it requires a vector-valued input $\pmb{\eta}$) and integrating $E_A(\pmb{\eta})$ over the polar angle of $\pmb{\eta}$, one finds
\begin{equation}
    T_\mathrm{PB,2}(\theta_1,\theta_2,\theta_3;\theta_4)=\frac{2\pi}{A} \bar{\mathcal{M}}_\mathrm{ap}^{2,0}(\theta_2,\theta_3;0)\int_0^\infty \dd\eta \; E_A(\eta)  \, \bar{\mathcal{M}}_\mathrm{ap}^{3,1}(\theta_1;\theta_4,\theta_4;\eta) + 2\, \mathrm{Permutations\,of}\,(\theta_1,\theta_2,\theta_3)\;,
    \label{eq: T_2 final}
\end{equation}
where for a survey of square area, $E_A(\eta)$ is given in Appendix A.2 of \cite{Heydenreich2020}\footnote{For the case $\sqrt{A}\leq\eta\leq\sqrt{2A}$ Eq. A.11 in \cite{Heydenreich2020} has a factor 2 typo.} as
\begin{equation}
    E_A(\eta)=\begin{cases}
        \frac{1}{A\pi}\left[A\pi-\left(4\sqrt{A}-\eta\right)\,\eta\right]\quad & \text{for}\quad\eta\leq \sqrt{A}\;,\\
        \frac{2}{\pi}\left[2\sqrt{\frac{\eta^2}{A}-1}-1-\frac{\eta^2}{2A}-\arccos\left(\frac{\sqrt{A}}{\eta}\right)+\arcsin\left(\frac{\sqrt{A}}{\eta}\right)\right]\quad & \text{for}\quad\sqrt{A}\leq\eta\leq\sqrt{2A}\;,\\
        0 \quad & \text{for}\quad \sqrt{2A}\leq\eta\;.
    \end{cases}
\end{equation}
Following the same procedure for $T_\mathrm{PB,1}$ and $T_\mathrm{PB,3}$ leads to
\begin{align}
    T_\mathrm{PB,1}(\theta_1,\theta_2,\theta_3;\theta_4) =&\;\mathcal{M}_\mathrm{ap}^{3,0}(\theta_1,\theta_2,\theta_3;\pmb{0}) \, \mathcal{M}_\mathrm{ap}^{2,0}(\theta_4,\theta_4;\pmb{0}) \nonumber\\ 
    =&\; \bar{\mathcal{M}}_\mathrm{ap}^{3,0}(\theta_1,\theta_2,\theta_3;0) \, \bar{\mathcal{M}}_\mathrm{ap}^{2,0}(\theta_4,\theta_4;0)\;,
    \label{eq: T_1 second approach} \\
    T_\mathrm{PB,3}(\theta_1,\theta_2,\theta_3;\theta_4) =&\;\frac{2}{A} \int \dd^2\eta \; E_A(\pmb{\eta}) \, \mathcal{M}_\mathrm{ap}^{3,2}(\theta_2,\theta_3;\theta_4;\pmb{\eta}) \, \mathcal{M}_\mathrm{ap}^{2,1}(\theta_1;\theta_4;\pmb{\eta}) +  2\, \mathrm{Permutations\,of}\,(\theta_1,\theta_2,\theta_3)\nonumber\\
    =&\;\frac{4\pi}{A} \int_0^\infty \dd\eta \; E_A(\eta) \, \bar{\mathcal{M}}_\mathrm{ap}^{3,2}(\theta_2,\theta_3;\theta_4;\eta) \, \bar{\mathcal{M}}_\mathrm{ap}^{2,1}(\theta_1;\theta_4;\eta) +  2\, \mathrm{Permutations\,of}\,(\theta_1,\theta_2,\theta_3)\;.
    \label{eq: T_3 second approach}
\end{align}
We note that because Eq.~\eqref{eq: cov cumulants} is written in cumulants, $T_\mathrm{P5}$ cannot be simply connected to the fifth-order correlation function. Therefore, to calculate $T_\mathrm{P5}$ in terms of correlation functions, we subtract $T_\mathrm{PB,1}$, $T_\mathrm{PB,2}$, and $T_\mathrm{PB,3}$ as defined above, such that
\begin{align}
    T_\mathrm{P5}(\theta_1,\theta_2,\theta_3;\theta_4)=&\;\frac{1}{A} \int \dd^2\eta \; E_A(\pmb{\eta}) \, \mathcal{M}_\mathrm{ap}^{5,3}(\theta_1,\theta_2,\theta_3;\theta_4,\theta_4;\pmb{\eta}) \nonumber \\ &- \, T_\mathrm{PB,1}(\theta_1,\theta_2,\theta_3;\theta_4)  - T_\mathrm{PB,2}(\theta_1,\theta_2,\theta_3;\theta_4) - T_\mathrm{PB,3}(\theta_1,\theta_2,\theta_3;\theta_4) \nonumber\\
    =&\;\frac{2\pi}{A} \int_0^\infty \dd\eta \; E_A(\eta) \, \bar{\mathcal{M}}_\mathrm{ap}^{5,3}(\theta_1,\theta_2,\theta_3;\theta_4,\theta_4;\eta) \nonumber\\
    &- \, T_\mathrm{PB,1}(\theta_1,\theta_2,\theta_3;\theta_4)  - T_\mathrm{PB,2}(\theta_1,\theta_2,\theta_3;\theta_4) - T_\mathrm{PB,3}(\theta_1,\theta_2,\theta_3;\theta_4)\;.\label{eq: T5 second approach}
\end{align}

\section{Analytical model validation}
\label{Sect:model validation}

With a model for the analytical cross-covariance of the second- and third-order aperture mass statistics in hand, it is important to test its consistency with $N$-body simulations.

\subsection{Covariance estimation methods from simulations}
A numerical cross-covariance can be obtained from simulations in two ways. The first and most straightforward method involves taking the unbiased sample covariance from multiple simulation realisations. For this approach, a significantly larger number of realisations (lines of sight) is required than the number of components in the data vector, and these realisations must be independent of one another. As discussed in Sect.~\ref{sec: cross-cov}, we need to exclude the boundaries of the fields by cutting off four times the largest aperture radius, $\theta_\mathrm{max}$. After spatially averaging the second- and third-order aperture mass over its centre coordinate $\pmb{\phi}$ ($\hat{\mathcal{M}}_\mathrm{ap}^2$ and $\hat{\mathcal{M}}_\mathrm{ap}^3$, respectively) for each realisation $i$, the sample covariance is computed using
\begin{equation}
    C^{\mathrm{sim}}_{\hat{\mathcal{M}}_\mathrm{ap}^{3,2}} (\theta_1,\theta_2,\theta_3;\theta_4)=\frac{1}{N-1}\sum^N_{i=1} \left[ \hat{\mathcal{M}}_\mathrm{ap}^3(\theta_1,\theta_2,\theta_3) \right]_i \,\left[ \hat{\mathcal{M}}_\mathrm{ap}^2(\theta_4) \right]_i - \frac{1}{N(N-1)}\sum^N_{i=1} \left[ \hat{\mathcal{M}}_\mathrm{ap}^3(\theta_1,\theta_2,\theta_3) \right]_i \, \sum^N_{j=1} \left[ \hat{\mathcal{M}}_\mathrm{ap}^2(\theta_4) \right]_j\;,
\end{equation}
where the sums are taken over the realisations, and $N$ represents the number of realisations.

The second method involves using aperture mass correlation function maps, as outlined in Sect.~\ref{sec: corrf}. It is important to note that the same boundary exclusion must be applied for this method. The advantage of this approach is that it allows us to obtain the $T$-terms separately.

The sum of $T_\mathrm{PB,2}$, $T_\mathrm{PB,3}$, and $T_\mathrm{P5}$, computed from the aperture mass correlation function maps (Eqs.~\ref{eq: T_2 final}, \ref{eq: T_3 second approach}, and \ref{eq: T5 second approach}), reproduces the full sample covariance. We use the individually measured $T$-terms solely for comparison with their analytical counterparts. In analyses that rely on the full numerical covariance, such as the cosmological parameter estimations in Sect.~\ref{sec: Cosmological Param}, we adopt the sample covariance as our reference, as this allows us to estimate errors via bootstrapping.

\subsection{Validation simulation - SLICS}
We use the Scinet-LIghtCone Simulations \citep[SLICS hereafter;][]{HarnoisDeraps2015,HarnoisDeraps2018} to validate the analytical model. These are dark matter only $N$-body simulations from which we use both shear catalogues and convergence maps.
The SLICS are a suite of ray-tracing simulations which use $1536^3$ particles placed in a $505\,h^{-1}\,\mathrm{Mpc}$ box. A flat $\mathrm{\Lambda CDM}$ cosmology is assumed, with normalised Hubble constant $h=0.69$, the normalisation of the power spectrum $\sigma_\mathrm{8}=0.83$, the equation of state of dark matter $w=-1$, matter density parameter $\Omega_\mathrm{m}=0.29$, baryon density parameter $\Omega_\mathrm{b}=0.047$, and primordial power spectrum spectral index $n_\mathrm{s}=0.969$.
For the set of convergence maps, we use $715$ independent lines of sight, all having an area of $10\times10\, \mathrm{deg^2}$, placed on a $7745\times7745$ pixel grid. All maps have a redshift of $z=0.897$, assume the flat-sky approximation, and do not contain shape noise. The aperture mass maps are calculated using the first equation in Eq.~\eqref{eq: map def}.
The shear catalogue set contains $927$ independent lines of sight, and has a galaxy number density
\begin{equation}
    n(z)\propto z^2\,\mathrm{e}^{-\left(z/z_0\right)^\beta}\;,
\end{equation}
where $z_\mathrm{0}=0.637$, $\beta=1.5$, and the galaxy density is normalised to $30\, \mathrm{arcmin^{-2}}$, mimicking a density that is expected in the Euclid survey (\citealt{laureijs2011euclid}; \citetalias{euclid2025}). The galaxies have a maximum redshift of $z_\mathrm{max}=3$. Each realisation has again an area of $10\times10\, \mathrm{deg^2}$. Every galaxy contains a shear, upon which an intrinsic ellipticity is added as shape noise. This is done by adding a random ellipticity to the shear from a Gaussian with a two-component ellipticity dispersion of $\sigma_\mathrm{\epsilon}^2=(0.37)^2$. The aperture mass is estimated by a weighted summation over the ellipticities of the galaxies
\begin{equation}
    \Hat{\mathcal{M}}_\mathrm{ap}(\pmb{\phi},\theta)=\frac{1}{n_\mathrm{gal}(\pmb{\phi})}\,\sum_i^N Q_\theta(\abs{\pmb{\phi}-\pmb{\vartheta}_i})\,\epsilon_{\mathrm{t},i}\;,\quad \text{with}\quad n_\mathrm{gal}(\pmb{\phi})=\sum_i^N Q_\theta(\abs{\pmb{\phi}-\pmb{\vartheta}_i})\;,
\end{equation}
where $N$ is the total number of galaxies in the catalogue and $\epsilon_{\mathrm{t},i}$ is the sum of the tangential shear plus the tangential intrinsic ellipticity for the galaxy $i$. This is a discrete approximation of the second definition of the aperture mass in Eq.~\eqref{eq: map def}. We test all filter combinations of $\theta\in \{4\arcmin,8\arcmin,16\arcmin\}$. The aperture centres of the convergence maps and shear catalogues are placed in a region of $7.87\times7.87\, \mathrm{deg^2}$ after boundary removal.

\subsection{Results}
\label{sec: results}
To compute the analytical cross-covariance using Eqs. (\ref{eq: T2, (3)}--\ref{eq: T4}), we require the convergence polyspectra. These are obtained by substituting the underlying matter polyspectra: the matter power spectrum is taken from the revised \texttt{Halofit} \citep{Takahashi2012}, and the bispectrum from \texttt{BiHalofit} \citep{Takahashi2020}. The tetraspectrum is approximated using the halo model, following the formalism of \citet{Cooray2002}, and detailed in Appendix~\ref{app: tetra}. Each polyspectrum is then projected into convergence space via Limber integration (Eq.~\ref{eq: limber}) to obtain the polyspectra for the convergence.

\subsubsection{Validity of the large-field approximation}
\label{sec: result large}
\begin{figure*}
    \centering
    \begin{subfigure}{0.465\linewidth}
        \includegraphics[width=\linewidth]{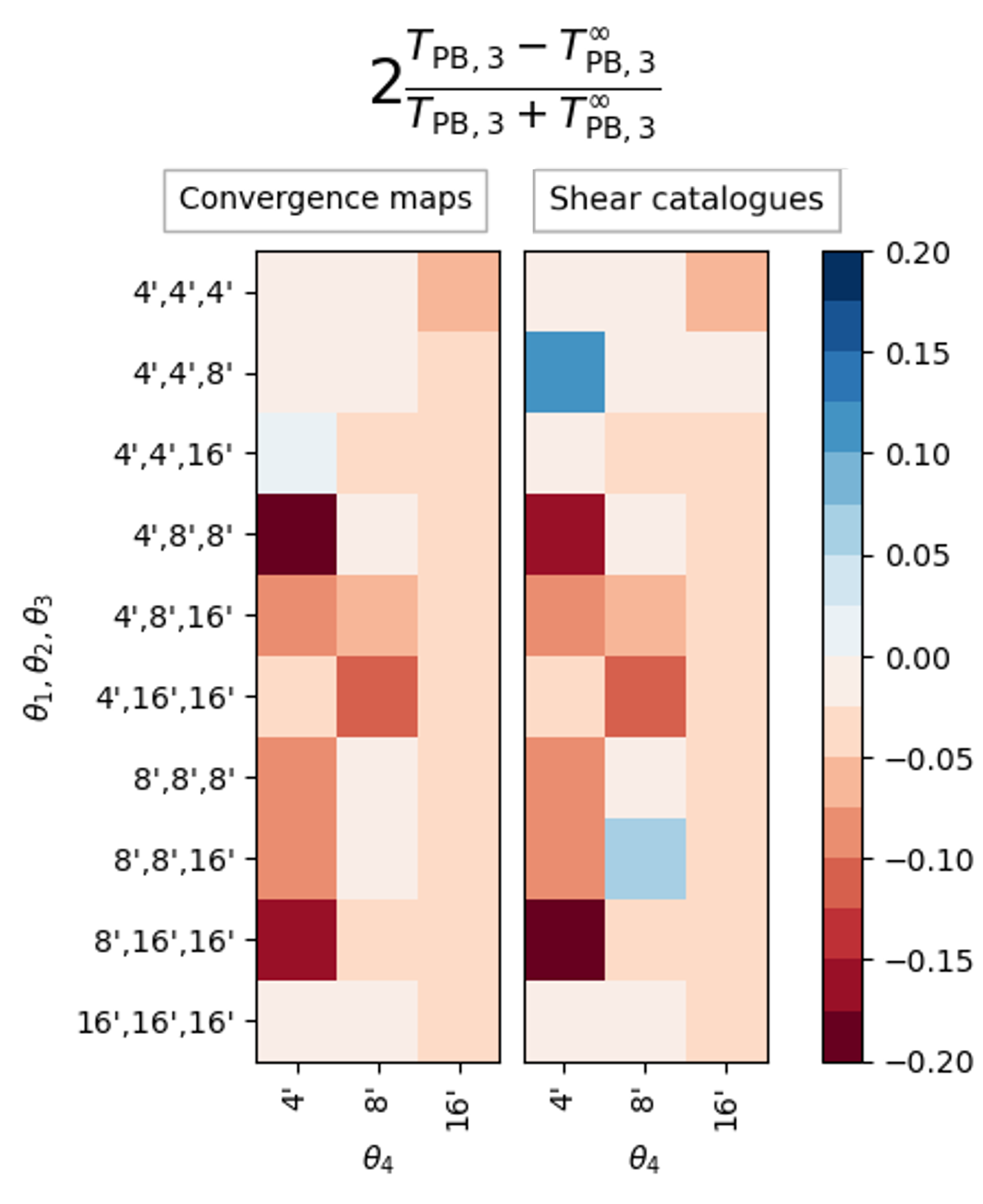}
    \end{subfigure}
    \hfill
    \begin{subfigure}{0.465\linewidth}
        \includegraphics[width=\linewidth]{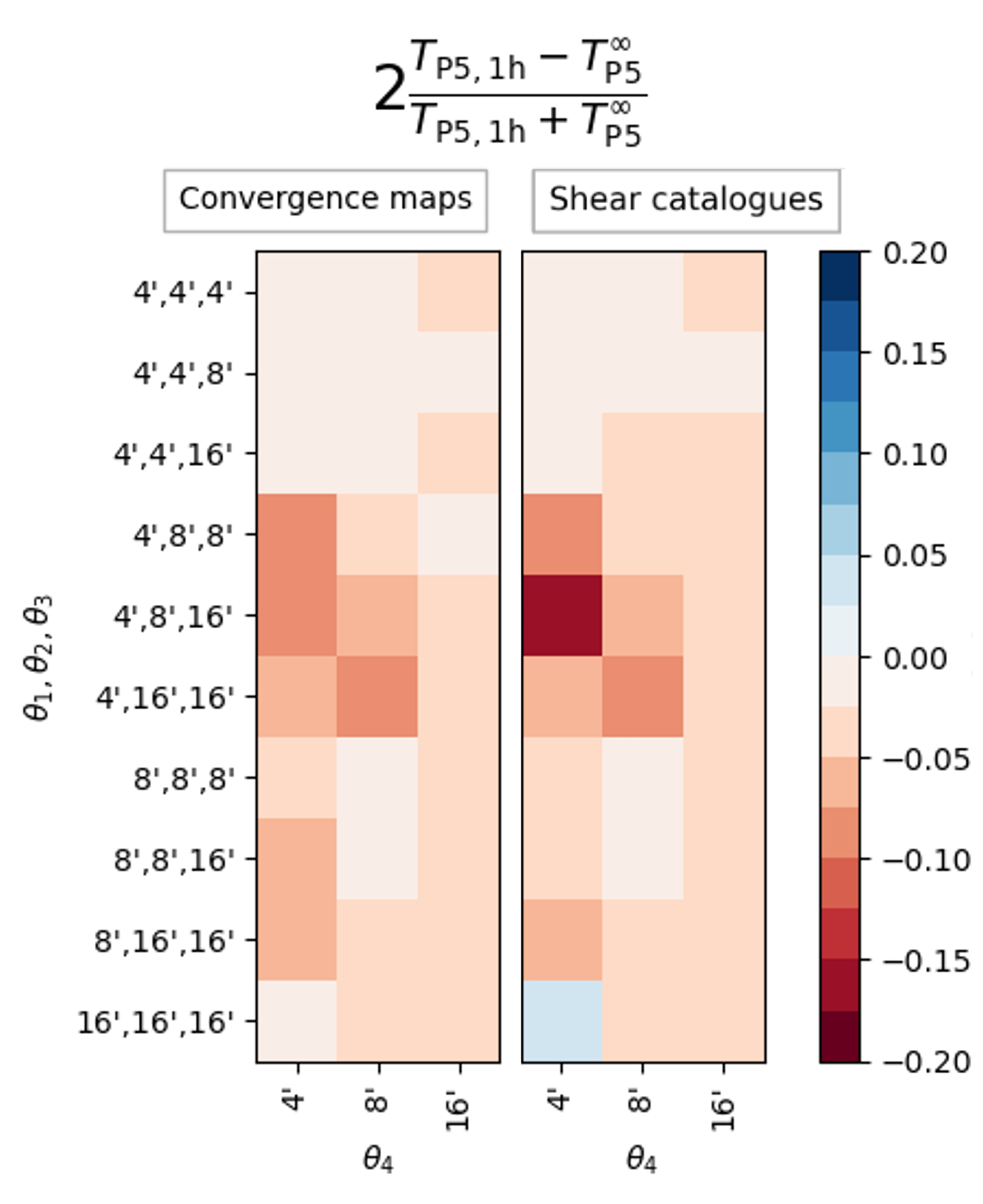}
    \end{subfigure}
    \\
    \begin{subfigure}{0.53\linewidth}
        \includegraphics[width=\linewidth]{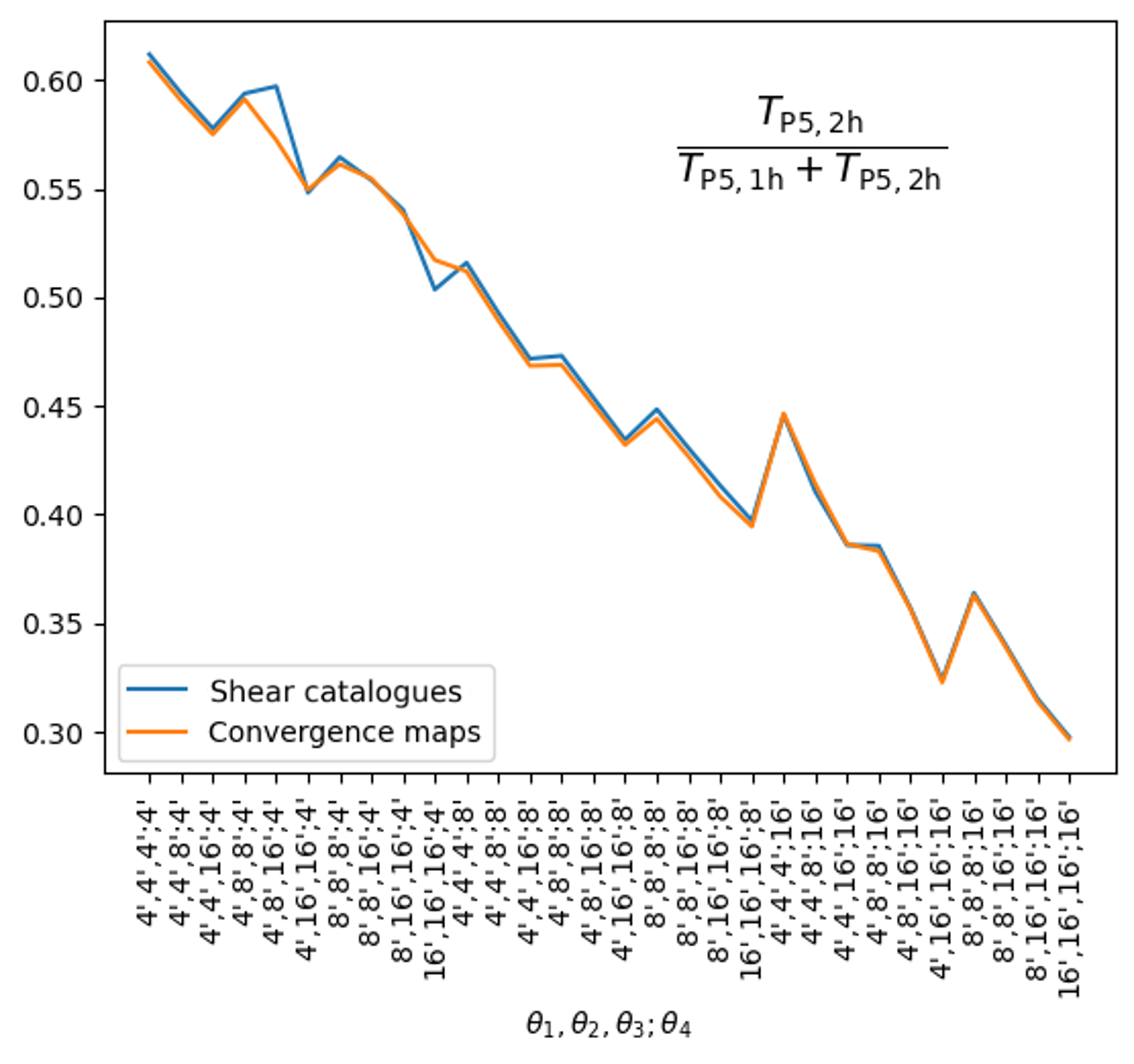}
    \end{subfigure}
    \hfill
    \begin{subfigure}{0.465\linewidth}
        \includegraphics[width=\linewidth]{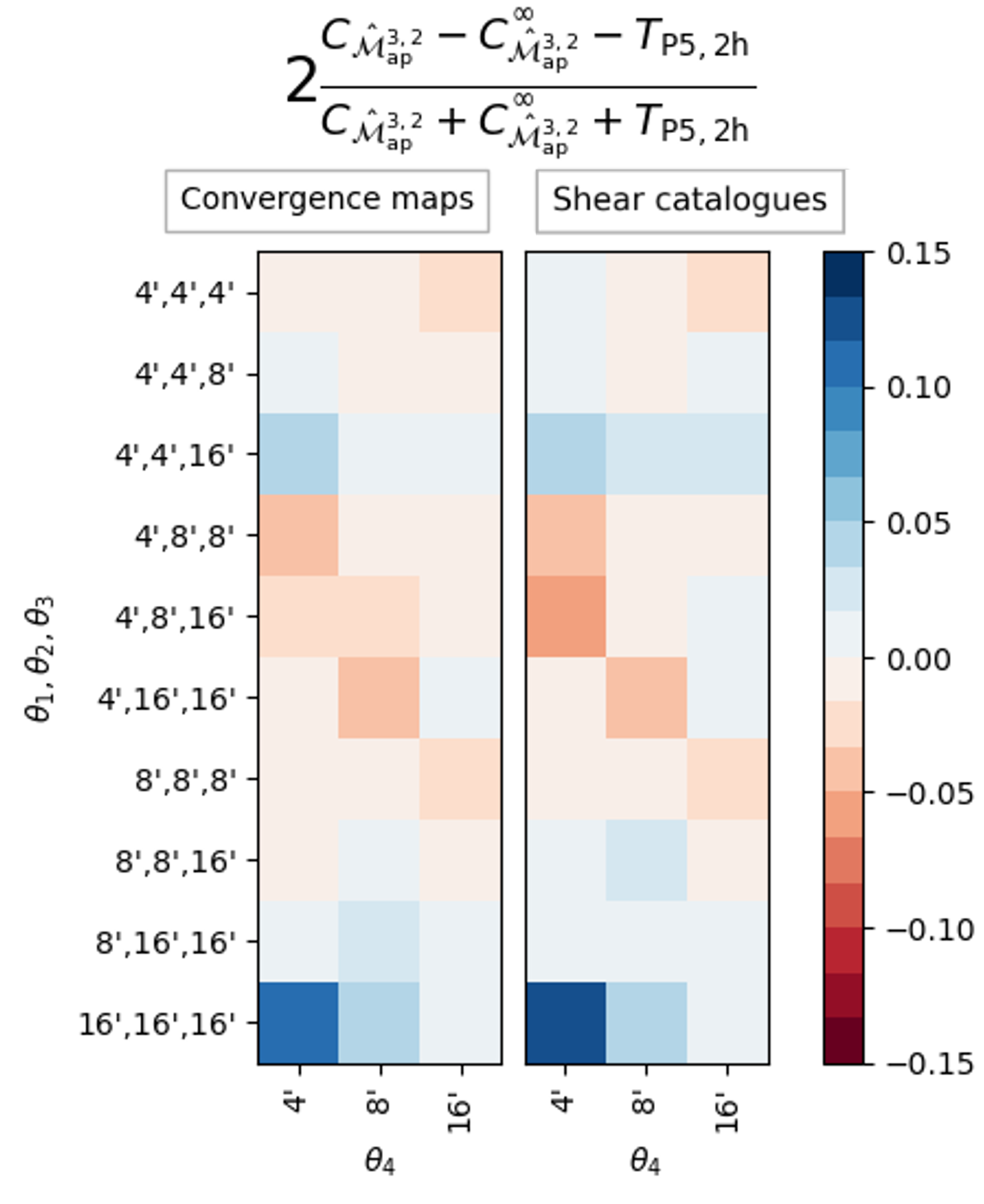}
    \end{subfigure}
     \caption{\textit{Top left}: Fractional difference between $T_\mathrm{PB,3}$ and $T_\mathrm{PB,3}^\infty$ for the convergence maps and the shear catalogues with shape noise. \textit{Top right}: Fractional difference between $T_\mathrm{P5,1h}$ and $T_\mathrm{P5}^\infty$ for the convergence maps and the shear catalogues. $T_\mathrm{P5}^\infty$ does not include the two-halo term $T_\mathrm{P5,2h}$ as it is a finite-field term and therefore zero in the large-field approximation (Appendix~\ref{app: tetra}). \textit{Bottom left}: Relative contribution of the two-halo term $T_\mathrm{P5,2h}$ to the total $T_\mathrm{P5}$. The blue line for the shear catalogues and the orange line for the convergence maps of the SLICS. \textit{Bottom right}: Fractional difference between the analytical cross-covariance and its large-field approximation plus the two-halo term $T_\mathrm{P5,2h}$ for both datasets.}
    \label{fig: inf terms}
\end{figure*}

First, we compare the analytical terms $T_\mathrm{PB,3}$ and $T_\mathrm{P5,1h}$ with their large-field approximation as shown in the top panels of Fig.~\ref{fig: inf terms}.
$T_\mathrm{PB,3}$ agrees with its large-field approximation within $5\%$ for $\sim 70\%$ of the filter combinations, both for the convergence maps without shape noise and the shear catalogues with shape noise. The larger deviations ($\sim 20\%$) are present mostly for small aperture radii combinations (combinations with $ 4\arcmin$). It seems larger aperture radii converge to the large-field approximation more rapidly than smaller ones. The large-field approximation generally overpredicts $T_\mathrm{PB,3}$. For $T_\mathrm{P5,1h}$, we see similar results; the one-halo term and the large-field approximation agree within less than $5\%$ for $\sim 80\%$ of the filter combinations in both datasets. For the shear catalogues, the filter combination $(4\arcmin,8\arcmin,16\arcmin;4\arcmin)$ differs by about $17\%$. This larger discrepancy is probably due to the inaccuracy of the integration in $T_\mathrm{P5,1h}$. To calculate it, we have to perform a nine-dimensional integration (one dimension coming from the mass integral in Eq.~\eqref{eq: I terms} of the halo model and the other eight as seen in Eq.~\ref{eq: T4}), two of which have to integrate over the oscillating function $G_A$. This is computationally intensive, so we were restricted in calculating it more precisely. Also, for $T_\mathrm{P5,1h}$, we see that small filter radii have the largest deviation from the large-field approximation. It should be noted that above, we compared the one-halo term with the large-field approximation, as the two-halo term $T_\mathrm{P5,2h}$ is zero in this approximation (see Appendix~\ref{app: tetra}). In the bottom left panel of Fig.~\ref{fig: inf terms}, we show the relative contribution of $T_\mathrm{P5,2h}$ to the total $T_\mathrm{P5}=T_\mathrm{P5,1h}+T_\mathrm{P5,2h}$.
For both datasets, $T_\mathrm{P5,2h}$ makes up around $60\%$ of $T_\mathrm{P5}$ for small filter radii combinations, and for larger radii combinations, we see a minimal contribution of $30\%$. Although $T_\mathrm{P5,2h}$ rapidly decreases with field size, at the field size of the SLICS simulations, the contribution is still significant. So, whereas approximating $T_\mathrm{P5,1h}$ by $T^\infty_\mathrm{P5}$ does not introduce a significant error, the large-field approximation is not valid for $T_\mathrm{P5}$ on SLICS-sized fields, and $T_\mathrm{P5,2h}$ still has to be added to find a meaningful result.
The bottom right panel of Fig.~\ref{fig: inf terms} shows the fractional difference between the large-field approximation of the cross-covariance $C^\infty_{\Hat{\mathcal{M}}_\mathrm{ap}^{3,2}}=T_\mathrm{PB,3}^\infty+T_\mathrm{P5,1h}^\infty$ plus the two-halo term $T_\mathrm{P5,2h}$ and the cross-covariance $C_{\Hat{\mathcal{M}}_\mathrm{ap}^{3,2}}$. Almost all filter combinations show an agreement within $5\%$. The filter combination $(16\arcmin,16\arcmin,16\arcmin;4\arcmin)$ is the only one showing a larger discrepancy ($\sim 12\%$).

\subsubsection{Validation with $N$-body simulations}
\begin{figure*}
    \raggedright
    \includegraphics[width=0.49\linewidth]{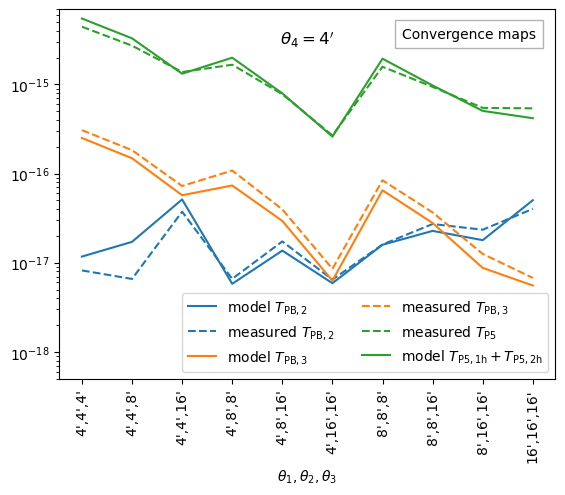}
     \includegraphics[width=0.49\linewidth]{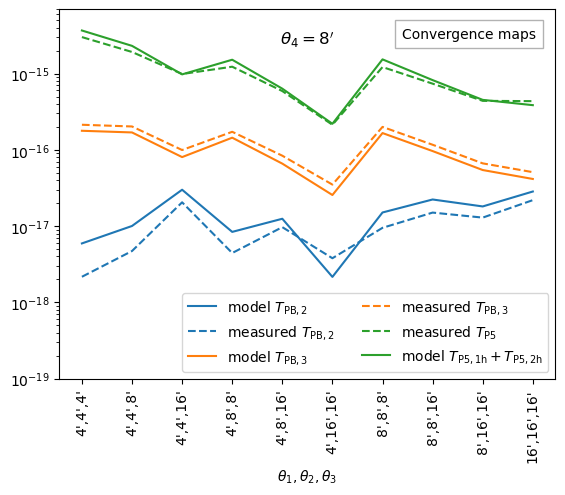}
     \includegraphics[width=0.49\linewidth]{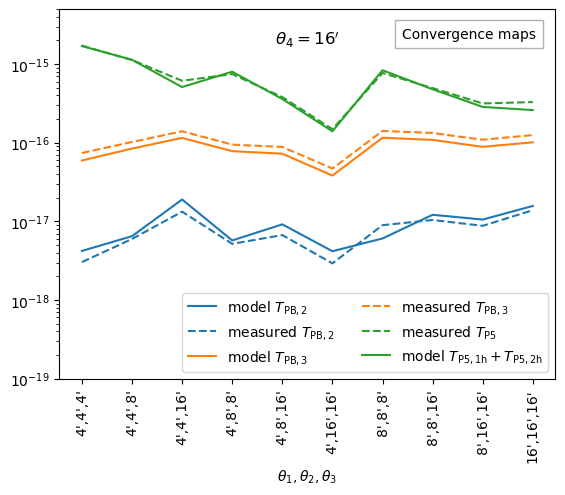}
     \includegraphics[height=0.43\linewidth]{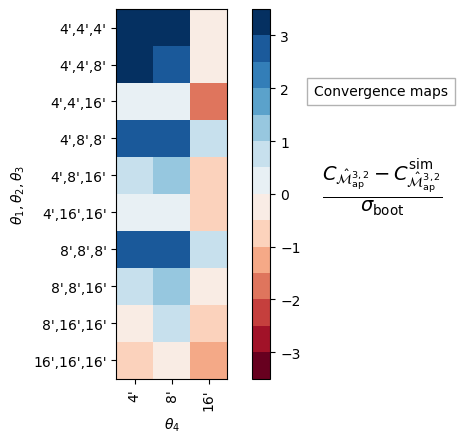}
      \caption{Individual terms of the model cross-covariance, plotted together with the numerically measured ones of the SLICS without shape noise (convergence maps). The top left plot displays the terms when fixing $\theta_4=4\arcmin$, the top right when fixing $\theta_4=8\arcmin$, and the bottom left fixing $\theta_4=16\arcmin$. The bottom right heat map shows the difference between the modelled cross-covariance and the sample covariance normalised by the bootstrapping error.}
    \label{fig: SLICS no noise}
\end{figure*}

\begin{figure*}
    \raggedright
    \includegraphics[width=0.49\linewidth]{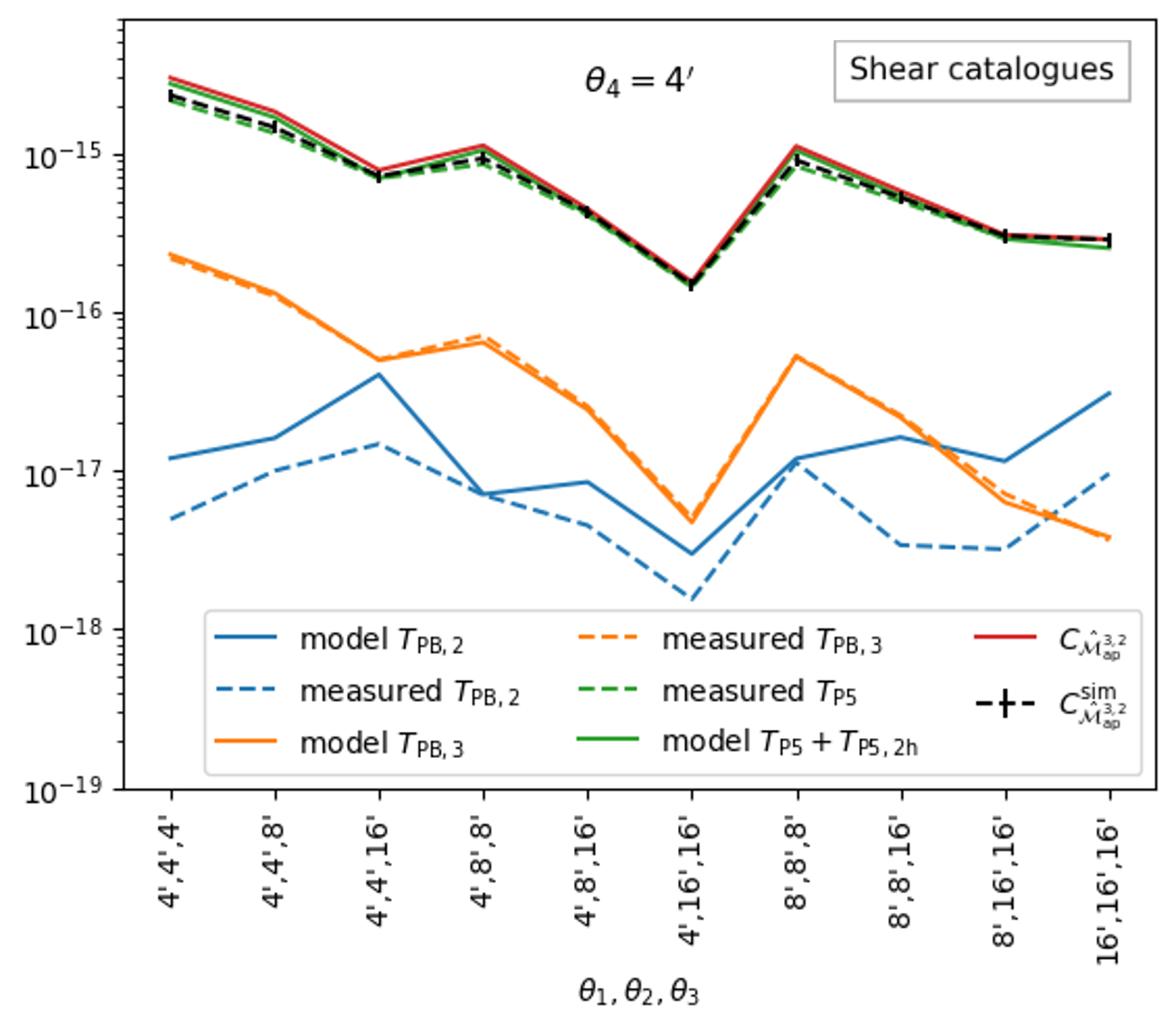}
     \includegraphics[width=0.49\linewidth]{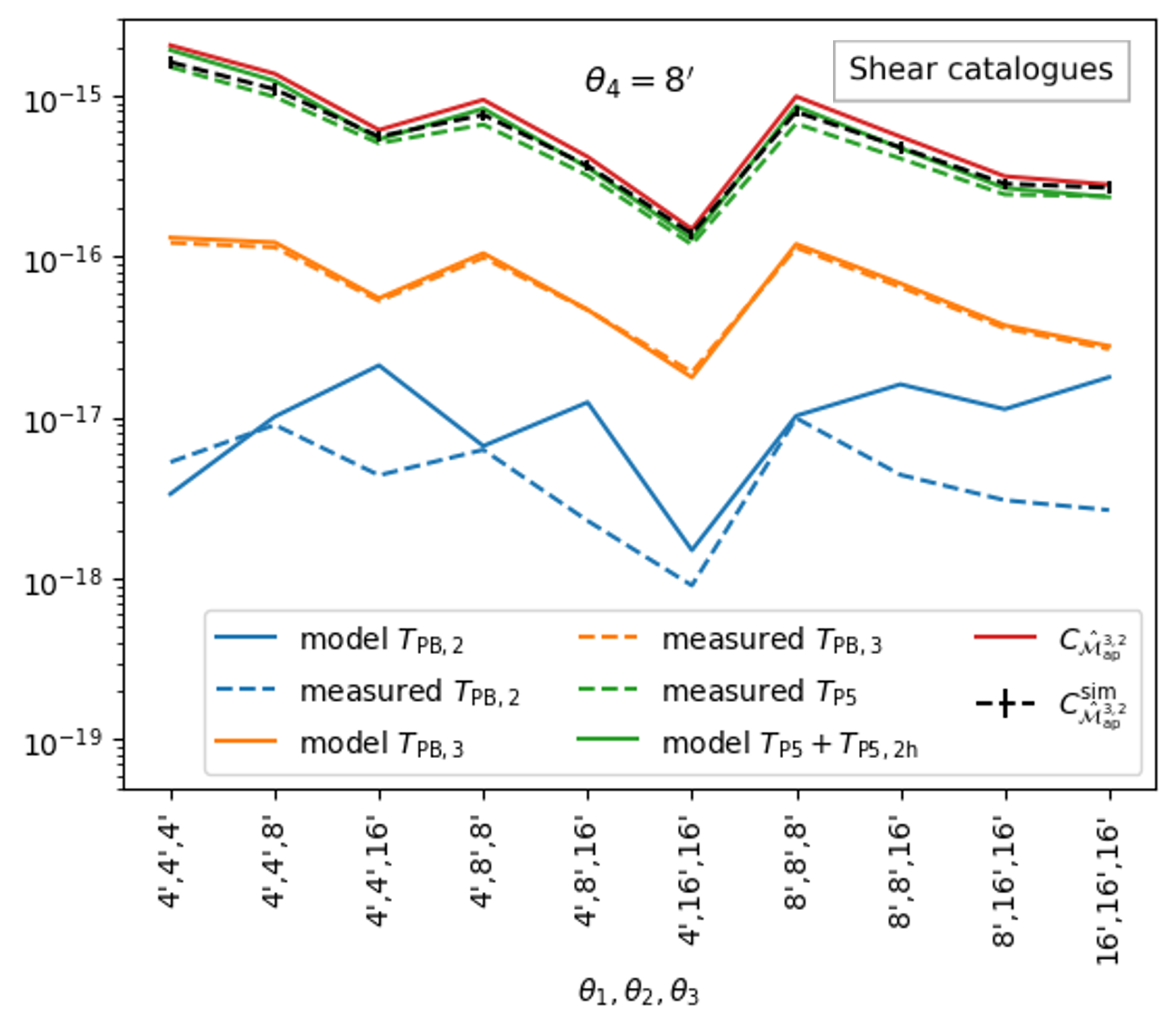}
     \includegraphics[width=0.49\linewidth]{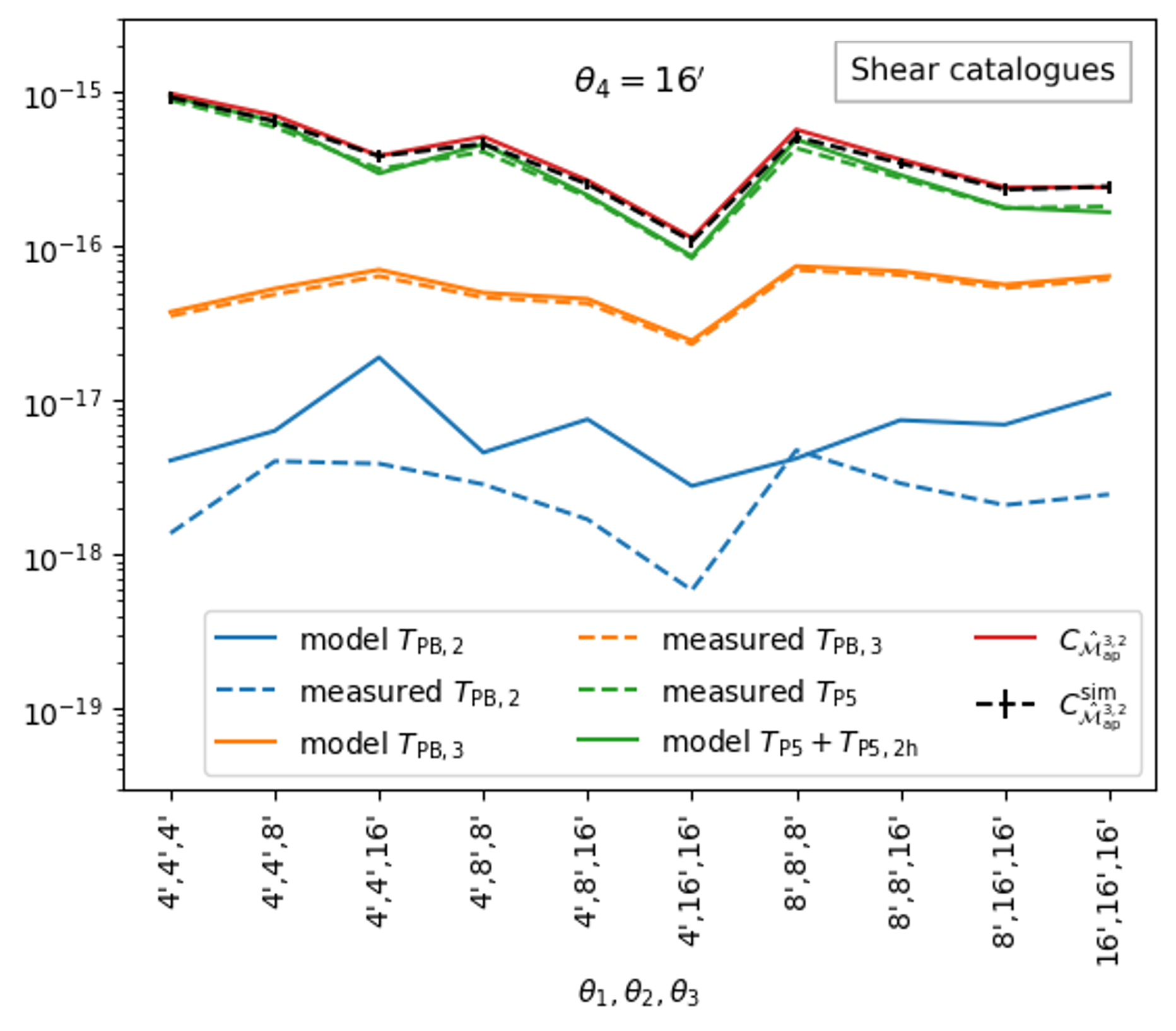}
     \includegraphics[height=0.43\linewidth]{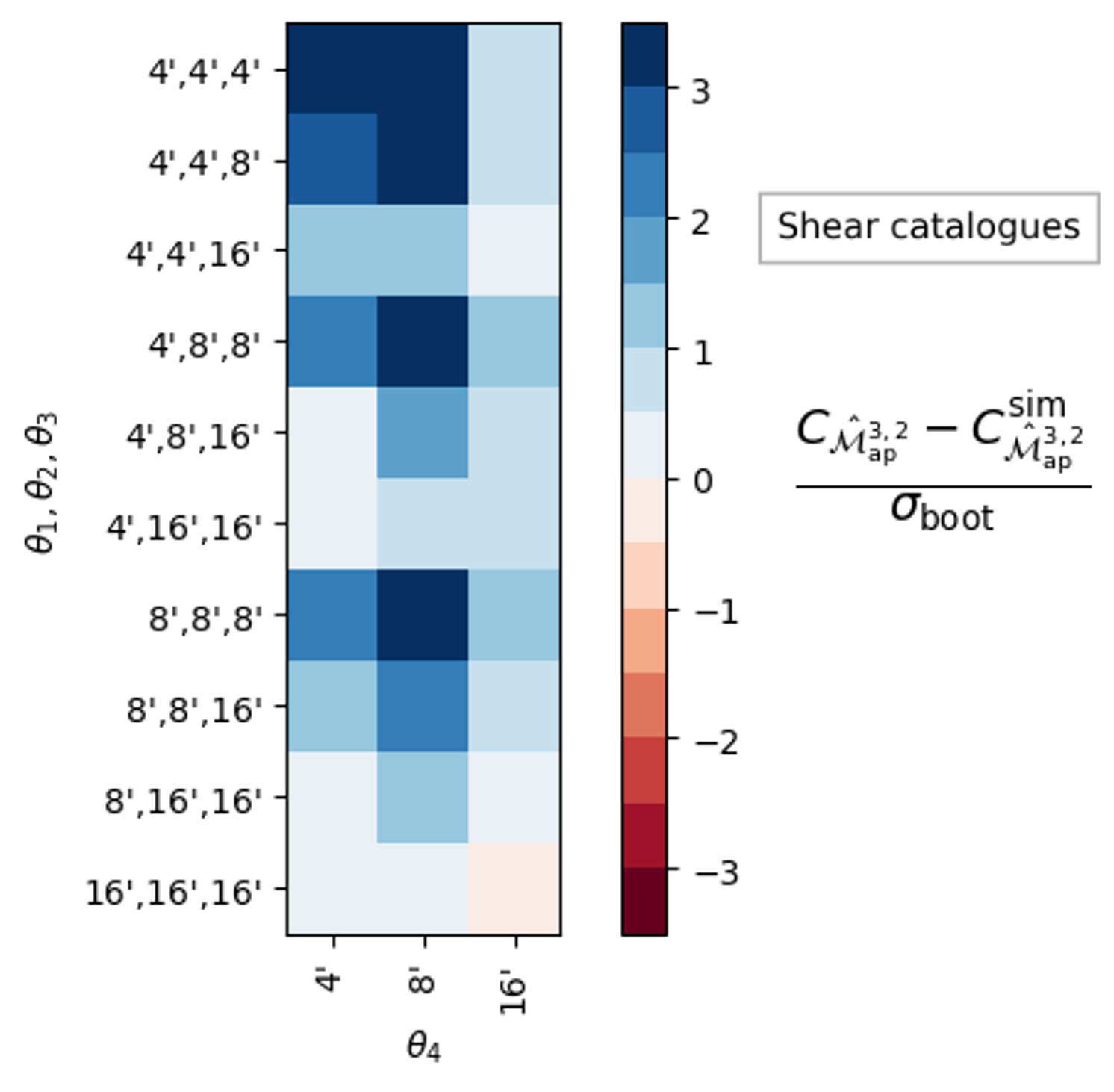}
      \caption{Individual terms of the analytical cross-covariance, plotted together with the numerically measured ones of the SLICS with shape noise (shear catalogues). The total analytical cross-covariance and the measured sample covariance are also plotted. The top left plot displays the terms when fixing $\theta_4=4\arcmin$, the top right when fixing $\theta_4=8\arcmin$, and the bottom left plot when fixing $\theta_4=16\arcmin$. The bottom right heat map shows the difference between the modelled cross-covariance and the sample covariance normalised by the bootstrapping error.}
    \label{fig: SLICS}
\end{figure*}

In Fig.~\ref{fig: SLICS no noise} and Fig.~\ref{fig: SLICS}, we compare the individual terms of the cross-covariance, both analytical and numerical, for the convergence maps and the shear catalogues, respectively. For all aperture radii combinations, $T_\mathrm{P5,1h}+T_\mathrm{P5,2h}$ (which depends on the tetraspectrum) dominates. Furthermore, we observe that for almost all combinations, $T_\mathrm{PB,2}$ is sub-dominant.

$T_\mathrm{P5}$ in general slightly over-predicts its numerical counterpart. The analytical $T_\mathrm{PB,3}$ is marginally smaller than the numerical one for the convergence maps. Conversely, for the shear catalogues, we see that the analytical $T_\mathrm{PB,3}$ is generally larger than the numerical one. The cause of the minor discrepancies between the numerical and analytical terms can be various. First, we expect the power-, bi-, and tetraspectrum to differ slightly in our model code and the SLICS simulation (see Appendix A in \citealp{Heydenreich2023} for an example of the discrepancy between the measured and modelled bispectrum of the \texttt{BiHalofit}). This can be due to the binning, finite resolution, smoothing of the simulation fields, and uncertainties in \texttt{Halofit} and \texttt{BiHalofit}. Also, we describe $T_\mathrm{P5}$ with the halo model, which is an approximate model. Moreover, we only include the most significant contribution of the two-halo term. Taking these discrepancies into account, it can be stated that the numerical and analytical terms agree reasonably well. Only for the SLICS with shape noise, we see that the modelled $T_\mathrm{PB,2}$ does not match the one measured from the simulation. This is probably due to the difficulties in measuring a finite-field term when noise is included. Taking the large-field approximation in Eq.~\eqref{eq: T_2 final} would mean neglecting $E_A(\pmb{\eta})$, after which the integral is, by definition, zero. That means the only reason it is non-zero for a finite field is the inclusion of $E_A(\pmb{\eta})$. This function is a slowly varying function that leads to an amplification of the shape noise in the measured correlation functions. We, therefore, do not expect an exact match between the analytical and measured $T_\mathrm{PB,2}$. Fortunately, as the finite-field term is sub-dominant, this has a negligible effect on the total cross-covariance. For the shear catalogues, we show the total analytical cross-covariance and the sample covariance, including error bars, in Fig.~\ref{fig: SLICS}. The model cross-covariance is just the sum of the separate terms. It is seen that the tetraspectrum term almost exclusively determines it.

Also shown in the heat maps in Fig.~\ref{fig: SLICS no noise} and Fig.~\ref{fig: SLICS} is the difference between the analytical cross-covariance and the sample covariance, normalised by the bootstrapping error. We observe that all filter combinations agree within $3.5\,\sigma_\mathrm{boot}$, with more than $50\%$ lying within the $1\,\sigma_\mathrm{boot}$ range. The model slightly over-predicts the sample covariance for most filter combinations, with the largest differences occurring for the smaller filter combinations. This is likely due to the differences mentioned earlier.

\section{Cosmological parameter estimation}
\label{sec: Cosmological Param}
\begin{figure*}
    \raggedright
    \begin{subfigure}{0.49\linewidth}
        \includegraphics[width=\linewidth]{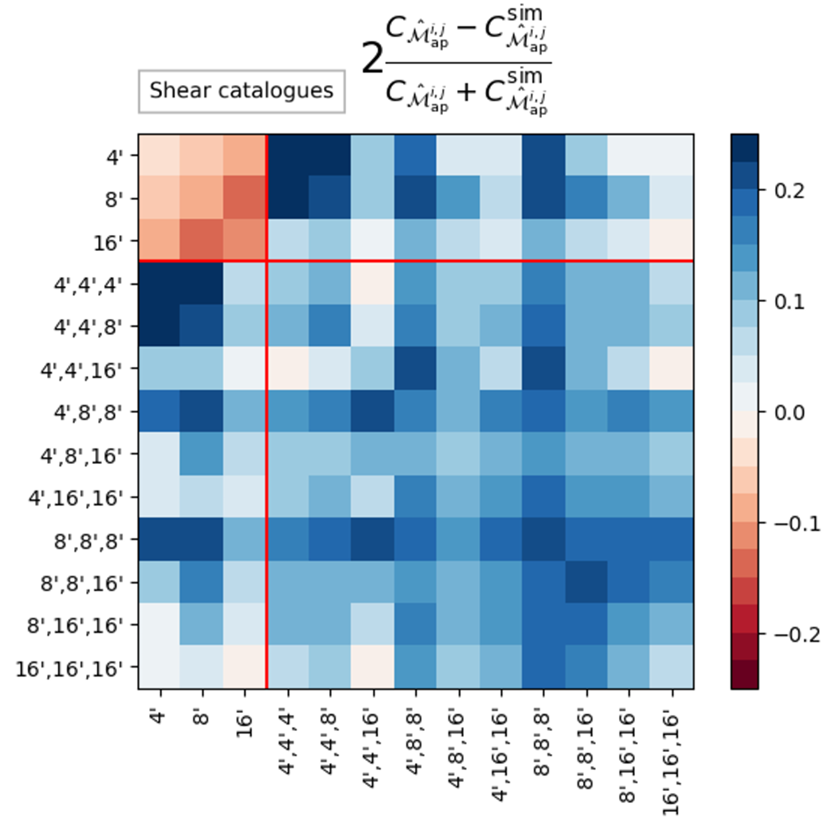}
    \end{subfigure}
    \hfill
    \begin{subfigure}{0.49\linewidth}
        \includegraphics[width=\linewidth]{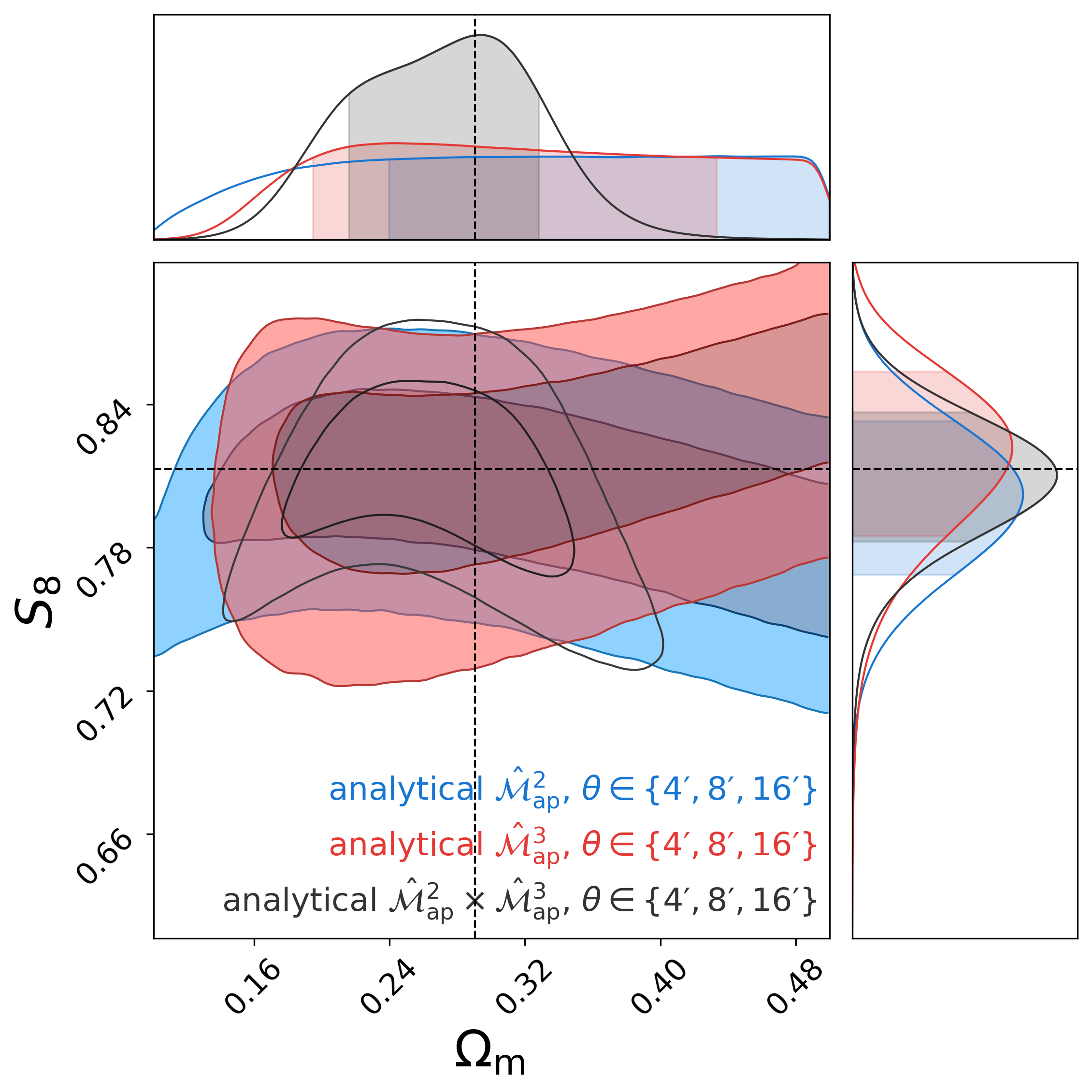}
    \end{subfigure}
     \caption{\textit{Left}: Fractional difference between the analytical and numerical joint covariance matrix of the second- and third-order aperture mass statistics. The red lines divide the parts of the joint covariance matrix. The $3\times3$ matrix at the top left is the covariance matrix of $\Hat{\mathcal{M}}_\mathrm{ap}^2$ calculated in \cite{linke2024supersample}. The $10\times10$ matrix at the bottom right is the covariance matrix of $\Hat{\mathcal{M}}_\mathrm{ap}^3$ from \citetalias{Linke2022b}. The non-square matrices at the top right and bottom left are the cross-covariance derived in this paper. \textit{Right}: Analytical $S_8$–$\Omega_\mathrm{m}$ contours from the second- and third-order aperture mass statistics, shown separately and in combination.}
    \label{fig: covariance matrix}
\end{figure*}

With an analytical cross-covariance of the second- and third-order aperture mass statistics now available, we can combine it with the analytical covariances of $\Hat{\mathcal{M}}_\mathrm{ap}^2$ and $\Hat{\mathcal{M}}_\mathrm{ap}^3$ to form a joint covariance matrix. The left panel of Fig.~\ref{fig: covariance matrix} illustrates the fractional difference between the analytical and numerical joint covariance matrices for the SLICS with shape noise (shear catalogues).
In this section, we compare the posterior distribution of $S_8$ and $\Omega_\mathrm{m}$ obtained from the analytical joint covariance with those derived from the SLICS simulation with shape noise. We use a Markov Chain Monte Carlo (MCMC) sampling method for this comparison. It should be noted that the purpose of this paper is not to provide realistic estimates for either the clustering or the matter density parameter. Rather, this section aims to demonstrate that similar posterior distributions can be obtained from the analytical model when compared to the numerical posteriors. The posterior distribution of the analytical model is proportional to
\begin{equation}
    \pmb{P}\left[\pmb{m}(\pmb{\Theta})\,|\,\pmb{d},\Tilde{C}\right] \propto |\Tilde{C}|^{-\frac{1}{2}} \,\mathrm{e}^{-\chi^2/2}\; ,
    \label{eq: posteriors ana}
\end{equation}
where $\pmb{m}$ is the model vector containing the predicted values of $\expval{\mathcal{M}_\mathrm{ap}}^2(\theta_i)$ and $\expval{\mathcal{M}_\mathrm{ap}}^3(\theta_j,\theta_k,\theta_l)$ given the cosmological parameters $\pmb{\Theta}$; $\pmb{d}$ is the data vector, containing the measured second- and third-order aperture mass statistics; and $\Tilde{C}$ is the corresponding covariance matrix.
For the numerical parameter estimation, we need to consider that we estimated the covariance from a finite sample. A correction for this is described in \cite{Percival2021}. The numerical posterior distribution is proportional to
\begin{equation}
    \pmb{P}\left[\pmb{m}(\pmb{\Theta})\,|\,\pmb{d},\Tilde{C}\right] \propto |\Tilde{C}|^{-\frac{1}{2}} \left( 1 + \frac{\chi^2}{n_\mathrm{r}-1}\right)^{-m/2}\; ,
    \label{eq:t_distribution}
\end{equation}
where $n_\mathrm{r}$ is the number of realisations. $m$ is a power-law index
\begin{equation}
    m = n_\theta+2+\frac{n_\mathrm{r}-1+B\,(n_\mathrm{d}-n_\theta)}{1+B\,(n_\mathrm{d}-n_\theta)}\;, \quad \text{with}\quad B=\frac{n_\mathrm{r}-n_\mathrm{d}-2}{(n_\mathrm{r}-n_\mathrm{d}-1)\,(n_\mathrm{r}-n_\mathrm{d}-4)}\;,
\end{equation}
$n_\theta$ being the number of parameters $\pmb{\Theta}$ and $n_\mathrm{d}$ the dimension of $\pmb{d}$. The $\chi^2$-distribution in Eqs.~\eqref{eq: posteriors ana} and \eqref{eq:t_distribution} is calculated as
\begin{equation}
    \chi^2 =  \left[\pmb{m}(\pmb{\Theta})-\vb{d}\right]^\mathrm{T} \Tilde{C}^{-1} \left[\pmb{m}(\pmb{\Theta})-\pmb{d}\right] \; .
\label{eq:chi2}
\end{equation}

We make our comparison by substituting $\Tilde{C}$ by $C_\mathrm{joint}$ and $C^\mathrm{sim}_\mathrm{joint}$, which are the analytical and numerical joint covariances, respectively, into Eqs.~\eqref{eq: posteriors ana}, \eqref{eq:t_distribution}, and \eqref{eq:chi2}. The numerical covariance $C^\mathrm{sim}_\mathrm{joint}$ is a positive definite matrix as it is calculated from the sample covariance. Our analytic covariance, as calculated to match the SLICS with shape noise, is not positive definite due to approximations and numerical errors, which poses a problem. Some of the subdominant eigenvalues of $C_\mathrm{joint}$ are found to be negative. This leads to a $\chi^2$-distribution with a non-zero probability of finding negative $\chi^2$ values, leading to incorrect posterior distributions.

To mitigate this problem, we approximate our joint analytic covariance with a principal component analysis (PCA). This method corresponds to performing an eigendecomposition of the covariance matrix and cutting away the subdominant (negative) eigenvalues and the corresponding eigenvectors. Because our covariance matrix is ill-conditioned, meaning the eigenvalues are close to zero, numerical instability can arise in the eigendecomposition, causing issues in the accuracy of the eigenvalues and eigenvectors. These instabilities arise due to finite precision in computer calculations, causing small perturbations or rounding errors. A numerically more stable method is the singular value decomposition (SVD). It is a generalisation to the eigendecomposition and is theoretically equivalent for symmetric matrices. Performing an SVD factorises our covariance matrix
\begin{equation}
    C_\mathrm{joint}=U\,\Sigma\,V^\mathrm{T}\;,
\end{equation}
where $\Sigma$ is a matrix with the singular values on the diagonal in descending order (which correspond to the absolute values of the eigenvalues for a symmetric matrix), and $U$ and $V^\mathrm{T}$ are real matrices with orthonormal columns and rows, respectively. The diagonal matrix scales coordinates, and the matrices $U$ and $V^\mathrm{T}$ represent rotations and reflections of the space. Up to different signs per column, $U$ and $V$ are equivalent to the matrix of eigenvectors of $C_\mathrm{joint}$ (but $U$ and $V$ are not the same; only the absolute values of the components are). We then cut away the diagonal terms corresponding to the negative eigenvalues and the corresponding columns and rows of $U$ and $V^\mathrm{T}$, respectively.
In our case, the original matrices are square $13\times13$ matrices. After the principal component analysis, $\Tilde{U}$ is a $13\times10$, $\Tilde{\Sigma}$ a $10\times10$, and $\Tilde{V}^\mathrm{T}$ a $10\times13$ matrix (as we had three negative eigenvalues), where the tildes represent the matrices after the PCA, such that
\begin{equation}
    \Tilde{C}_\mathrm{joint}=\Tilde{U}\,\Tilde{\Sigma}\,\Tilde{V}^\mathrm{T}\;.
    \label{eq: tilde cov}
\end{equation}
The new matrix $\Tilde{C}_\mathrm{joint}$ is a positive semi-definite matrix. As $\Tilde{C}_\mathrm{joint}$ is a rank 10 matrix with dimensions $13\times13$, it is singular. Hence, it is non-invertible. Taking the inverse of the separate terms on the right-hand side of Eq.~\eqref{eq: tilde cov} is thus not possible as $\Tilde{U}$ and $\Tilde{V}$ are not square matrices. Therefore, we take their pseudoinverse (denoted as a cross). As $\tilde{U}$ has orthonormal columns and $\tilde{V}^\mathrm{T}$ orthonormal rows, their pseudoinverse matrices are equal to their transpose matrices ($\Tilde{\Sigma}$ has an inverse, and thus the pseudoinverse is equal to the inverse)
\begin{equation}
    (\Tilde{U}\,\Tilde{\Sigma}\,\Tilde{V}^\mathrm{T})^+=(\Tilde{V}^\mathrm{T})^+\,\Tilde{\Sigma}^+\,\Tilde{U}^+=\Tilde{V}\,\Tilde{\Sigma}^{-1}\,\Tilde{U}^\mathrm{T}\;.
\end{equation}
Substituting this approximation into Eq.~\eqref{eq:chi2}, we find
\begin{equation}
    \chi^2 \approx  \left[\pmb{m}(\pmb{\Theta})-\vb{d}\right]^\mathrm{T} \Tilde{V}\,\Tilde{\Sigma}^{-1}\,\Tilde{U}^\mathrm{T} \left[\pmb{m}(\pmb{\Theta})-\pmb{d}\right] \; ,
\end{equation}
which has a strictly positive $\chi^2$-distribution. With this set-up, we perform a Metropolis-Hastings MCMC, trained on 6500 model points sampled in a Latin hypercube spanning the cosmological parameter space $\Omega_\mathrm{m}$ and $S_8$. We use an emulator adopting the \texttt{CosmoPower} framework from \citet{COSMOPOWER2022} to efficiently compute the model predictions $\expval{\hat{\mathcal{M}}_\mathrm{ap}}^2(\theta_i)$ and $\expval{\hat{\mathcal{M}}_\mathrm{ap}}^3(\theta_j,\theta_k,\theta_l)$ for any given parameter vector $\Theta$. The data vector $\pmb{d}$ is constructed from the emulator evaluated at the fiducial cosmological parameters corresponding to the SLICS simulations. All other cosmological parameters not varied in the analysis are fixed to their SLICS values.

\begin{figure}
     \centering
     \begin{subfigure}[b]{0.49\textwidth}
         \centering
         \includegraphics[width=\textwidth]{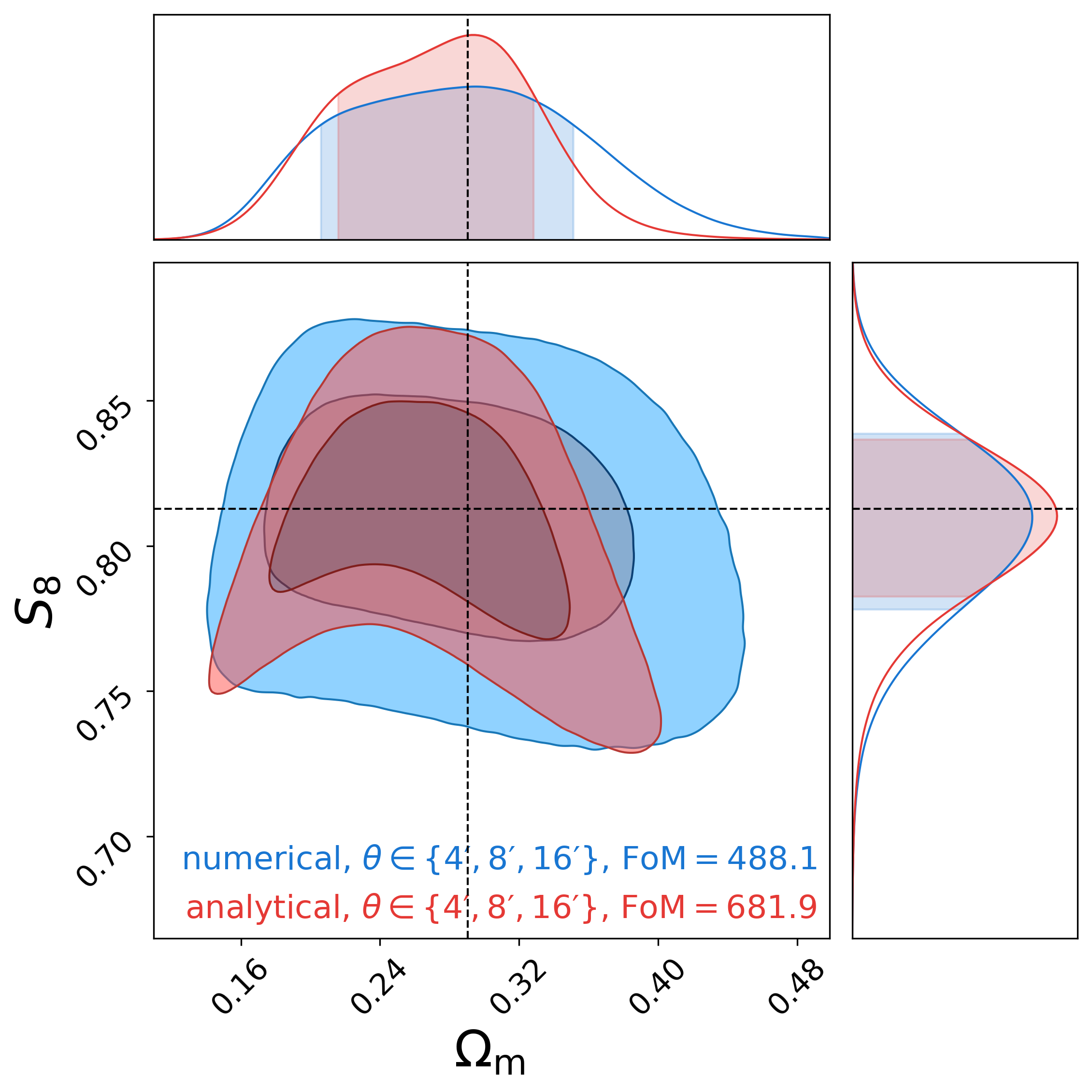}
         \label{fig:y equals x}
     \end{subfigure}
     \hfill
     \begin{subfigure}[b]{0.49\textwidth}
         \centering
         \includegraphics[width=\textwidth]{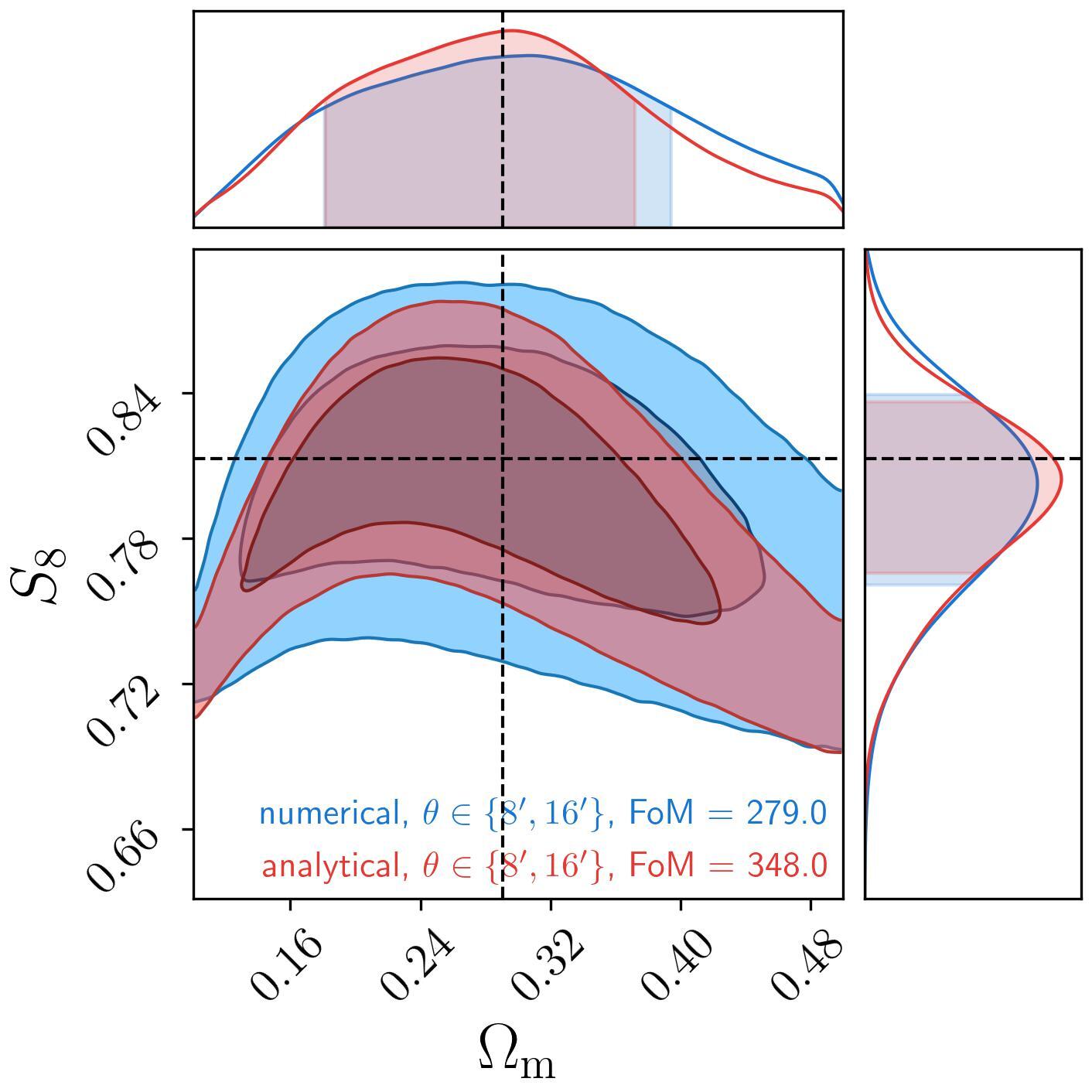}
         \label{fig:three sin x}
     \end{subfigure}
        \caption{Posterior contours of $S_8$ with $\Omega_\mathrm{m}$ obtained via MCMC. The blue contour is obtained via the sample covariance of the SLICS shear catalogues. The red contour is calculated via the analytical covariance. \textit{Left}: Using all the filter radii combinations in $\theta\in\{4\arcmin,8\arcmin,16\arcmin\}$. \textit{Right}: Using only the combinations of $\theta\in\{8\arcmin,16\arcmin\}$.}
        \label{fig: MCMC}
\end{figure}

The right panel of Fig.~\ref{fig: covariance matrix} shows the analytical posterior contours for $\hat{\mathcal{M}}_\mathrm{ap}^2$, $\hat{\mathcal{M}}_\mathrm{ap}^3$, and their joint posterior contour $\hat{\mathcal{M}}_\mathrm{ap}^2 \times \hat{\mathcal{M}}_\mathrm{ap}^3$. The degeneracy directions of the second- and third-order aperture mass statistics differ noticeably, resulting in significantly tighter constraints when the two are combined.
A comparison between the
analytical and numerical $S_8$-$\Omega_\mathrm{m}$ contours is shown in the left panel of Fig.~\ref{fig: MCMC}.
We calculate the figure of merit (FoM) with a parameter covariance $\Hat{C}$ obtained from the MCMC process as
\begin{equation}
    \mathrm{FoM}=\frac{1}{\sqrt{\mathrm{det}\Hat{C}}}\;.
\end{equation}
The fraction of the numerical over the analytical FoM is $0.72$. The analytical posteriors are noticeably smaller than the numerical ones. In the left panel of Fig.~\ref{fig: covariance matrix}, the largest discrepancies between the analytical and numerical cross-covariance occur at smaller filter radii scales (combinations with $4\arcmin$). Therefore, we also perform an MCMC with combinations of $8\arcmin$ and $16\arcmin$ only, as shown in the right panel of Fig.~\ref{fig: MCMC}. Excluding $4\arcmin$, $C_\mathrm{joint}$ does not contain any negative eigenvalues, so we did not perform a PCA. Here, the fraction of the numerical over the analytical FoM is $0.80$, which is an improvement compared to when all filter radii combinations are included.

\section{Discussion}
\label{sec: disc}
We derived an analytical model for the non-tomographic cross-covariance between the real-space estimators $\Hat{\mathcal{M}}_\mathrm{ap}^2$ and $\Hat{\mathcal{M}}_\mathrm{ap}^3$, and validated it against simulated data from SLICS, including both shape-noise-free convergence maps and shear catalogues with noise. The cross-covariance from the simulations was numerically determined through either the sample covariance or aperture mass correlation functions. For our selected set of aperture radii combinations, we found that the maximum deviation between the analytical and numerical results was $3.5$ standard deviations on the bootstrapping uncertainty, with over $70\%$ of the cases falling well within the two-sigma range. This was true for both datasets. The most likely cause of the discrepancies is a difference between the polyspectra used in the model and those from the simulation.

\subsection{The large-field approximation}
\label{sec: disc large}
We find that the large-field approximation has a negligible effect on the terms $T_\mathrm{P5,1h}$ and $T_\mathrm{PB,3}$. These can therefore be safely replaced by their large-field counterparts, $T_\mathrm{P5}^\infty$ and $T_\mathrm{PB,3}^\infty$, for a SLICS sized or larger field, if a reduction in computational cost is desired or if the integration over the survey window function $G_A$ leads to numerical instability.
The finite-field term $T_\mathrm{PB,2}$ is consistently subdominant across the range of filter combinations tested. Its omission has little impact on the overall cross-covariance and it can be neglected without significantly affecting the result.
In contrast, the finite-field term $T_\mathrm{P5,2h}$ is essential for achieving an accurate prediction. For combinations involving small aperture radii, it can contribute up to $60\%$ of the total $T_\mathrm{P5}$, which is the dominant component of the cross-covariance. Neglecting this term would therefore lead to a substantial underestimation of the covariance at small scales. Since the amplitude of finite-field corrections decreases more rapidly than linearly with increasing survey area, their relevance is expected to diminish in Stage IV surveys. In particular, for significantly larger survey fields than those of SLICS, the contribution from $T_\mathrm{PB,2}$ is negligible. Nonetheless, we recommend retaining $T_\mathrm{P5,2h}$ in any analysis that includes small-scale filters, even for larger-area surveys.

\subsection{Analytical vs. numerical cross-covariance and origins of inaccuracy}
\label{sec: disc analytical}
The analytical cross-covariance model presented in this work agrees well with the numerical results from simulations across a broad range of filter combinations. For the configurations tested, $50\%$ of the analytical predictions fall within the 1 $\sigma$ range of the bootstrap uncertainties, with the largest individual deviation reaching 3.5 $\sigma$. The agreement is notably better for combinations involving larger aperture radii ($16\arcmin$), while the largest discrepancies arise at small filter scales. Testing the model at even larger aperture radii, such as $32\arcmin$, in larger-volume simulations, for example, the full-sky GL simulations of \citet{Takahashi2017}, would provide valuable insight into the behaviour of the analytic cross-covariance at lower $\ell$ and help isolate contributions from different model components.

There are several possible reasons for the observed discrepancies at smaller filter radii. The most significant source of inaccuracy is most likely the halo model used to compute the tetraspectrum term $T_\mathrm{P5}$, which dominates the cross-covariance at all aperture scales. Among its components, the two-halo contribution $T_\mathrm{P5,2h}$ is particularly impactful at small radii. As no more accurate tetraspectrum model is currently available, this approximation remains necessary, though the inclusion of additional two-halo contributions may improve the model’s fidelity, especially for the small filter scales, where $T_\mathrm{P5,2h}$ is dominant. While the power spectrum and bispectrum models used to compute $T_\mathrm{PB,2}$ and $T_\mathrm{PB,3}$ also differ from those implicit in the simulations, these terms are subdominant, typically by one to two orders of magnitude, and thus contribute less to the overall discrepancy.

A secondary limitation arises from numerical integration accuracy. The calculation of $T_\mathrm{P5}$ involves high-dimensional integrals that are computationally expensive and sensitive to binning. Due to performance constraints, we employed relatively coarse binning in parts of the integration, which limits the achievable precision. Some of the residual discrepancies are likely due to this numerical limitation.

Several other approximations were used in the model but are unlikely to contribute significantly to the observed discrepancies. First, the flat sky approximation was used. This is a valid approximation, as the scales probed in this are below $\sim 2\degr$, where the induced error is negligible \citep{Kilbinger_2017}. Moreover, this approximation was also applied in the simulations, meaning it cannot account for any discrepancies. However, it is important to consider this approximation when extending to larger fields, where larger filter radii can be probed, as it is expected to introduce greater errors at those scales. Additionally, the Limber approximation was used. This approximation becomes unreliable on scales larger than $6\degr$ \citep{Deshpande_2020}. However, non-linear structure growth is minimal at these scales, meaning third-order statistics provide little additional information, making this limitation largely inconsequential.

In conclusion, the primary source of inaccuracy in the analytical cross-covariance model lies in the treatment of the tetraspectrum. Numerical integration precision contributes to a lesser extent, whereas other standard approximations, such as the flat-sky and Limber approximations, have a negligible impact on the angular scales considered.

\subsection{Analytic vs. numerical cosmological inference}
By combining the analytical cross-covariance derived here with the analytical $\Hat{\mathcal{M}}_\mathrm{ap}^3$ covariance of \citetalias{Linke2022b} and the $\Hat{\mathcal{M}}_\mathrm{ap}^2$ covariance of \cite{linke2024supersample}, we constructed a joint analytical covariance matrix for use in cosmological parameter inference. We then compared its performance to the numerically derived covariance from SLICS in constraining the parameters $\Omega_\mathrm{m}$ and $S_8$. 
The resulting constraints are broadly consistent between the two approaches. The analytical model yields slightly tighter confidence contours (Fig.~\ref{fig: MCMC}), indicating slightly optimistic constraining power. However, interpreting this difference requires care, as the overall agreement of the covariance components is not uniform. The fractional difference matrix in the left panel of Fig.~\ref{fig: covariance matrix} shows that the analytical  $\hat{\mathcal{M}}_\mathrm{ap}^2$ auto-covariance underpredicts its numerical counterpart, while both the cross-covariance and the $\hat{\mathcal{M}}_\mathrm{ap}^3$ auto-covariance tend to overpredict. This asymmetry suggests that the origin of the discrepancy in the contours is not easily attributable to a single term, such as the cross-covariance. In addition to the discrepancies for the cross-covariance mentioned before, a plausible explanation for the underpredicting $\hat{\mathcal{M}}_\mathrm{ap}^2$ auto-covariance lies in the modelling of the trispectrum. The analytical $\hat{\mathcal{M}}_\mathrm{ap}^2$ covariance includes a trispectrum term computed using only the one-halo contribution. If the trispectrum extracted from simulations (e.g. SLICS) is larger due to two-halo or higher-order contributions, this would lead to an underestimation of the variance in the analytical $\hat{\mathcal{M}}_\mathrm{ap}^2$ covariance, thereby shifting the contours accordingly. In contrast, the cross-covariance and $\hat{\mathcal{M}}_\mathrm{ap}^3$ covariance may slightly overpredict their counterparts due to different approximations or modelling assumptions.

In practice, the joint analytical covariance matrix is prone to numerical instability. Small-scale inaccuracies and inconsistencies among components can cause the matrix to become ill-conditioned, complicating operations such as matrix inversion. To mitigate this, PCA and SVD were applied to regularise the matrix. While PCA improves numerical stability, it also removes information from the data vector, leading to a slight degradation of the constraining power. These numerical issues highlight the sensitivity of cosmological inference to the stability and internal consistency of the covariance matrix, particularly when combining multiple orders of statistics. Future improvements to analytical models should aim not only for greater physical accuracy but also for improved numerical robustness.

\subsection{Further comments}
\label{sec: disc comments}
In realistic surveys, data often contain gaps due to masking, which prevents the direct computation of aperture mass statistics from convergence or shear maps using Eqs.~\eqref{eq: map def}. Aperture centres located near masked regions lead to weighted integrals over incomplete areas, introducing biases in the aperture mass. To address this, the estimators $\expval{\mathcal{M}_\mathrm{ap}^2}$ and $\expval{\mathcal{M}_\mathrm{ap}^3}$ should instead be derived from their relations to the two- and three-point shear correlation functions (2PCF and 3PCF), as proposed by \citet{Schneider_2005}. However, even if masking is not explicitly accounted for, \citetalias{Linke2022b} demonstrated that the analytical $\Hat{\mathcal{M}}_\mathrm{ap}^3$ covariance can still provide meaningful constraints by establishing upper and lower bounds. The upper bound is computed assuming a field size equal to the total unmasked survey area, while the lower bound excludes a boundary region equal to four times the largest aperture radius. For a SLICS-sized field, these bounds were found to be tight: the upper bound exceeded the lower bound by only $\sim 15\%$. In Stage IV surveys, the relative size of the excluded boundary will be smaller, implying even tighter bounds. Although we have not explicitly tested this for the analytical cross-covariance derived here, the underlying scaling arguments suggest that similar bounds should apply. This makes the analytical cross-covariance a practical tool even in the presence of realistic masking.

The analytical cross-covariance between $\Hat{\mathcal{M}}_\mathrm{ap}^2$ and $\Hat{\mathcal{M}}_\mathrm{ap}^3$ is closely related to the Fourier-space cross-covariance between the 3D power spectrum and bispectrum, as calculated by \citet{Sugiyama_2020}. While their formalism is three-dimensional, their approach closely parallels ours and can be translated to the 2D case via the Limber approximation. A direct comparison reveals that their expressions correspond to our large-field terms. Specifically, their Eq. (23) maps to our $T_\mathrm{PB,3}^\infty$ and their Eq. (26) corresponds to our $T_\mathrm{P5}^\infty$. This correspondence is evident from the structural similarity of their $k$-space with our $\ell$-space configurations. \citet{Sugiyama_2020} note that their expressions do not include super-sample covariance (SSC) terms, which explains the absence of certain contributions. In Fourier space, SSC arises from modes larger than the survey area and must be added explicitly. In contrast, real-space calculations naturally incorporate these large-scale modes within the measured survey window, eliminating the need for separate SSC corrections. A detailed discussion of this distinction, and the implicit inclusion of SSC in real-space statistics, is provided in \citet{linke2024supersample}.

\subsection{Future work}
\label{sec: disc future}
One clear avenue for extending this work is the incorporation of tomography. A tomographic approach provides significantly more information on the evolution of large-scale structure and is particularly effective at reducing degeneracies between cosmological parameters such as $\Omega_\mathrm{m}$ and $S_8$. It also enables meaningful constraints on a time-varying dark energy equation of state, $w(z)$. However, extending the current model to a tomographic framework is non-trivial. It leads to a substantial increase in the size of both the data vector and the covariance matrix, which increases computational cost and the risk of numerical instability. Analytical derivation of the tomographic covariance is, however, still more feasible than relying on simulations, given the substantial resources required to generate tomographic mock catalogues, but it introduces new challenges for validation. In particular, as simulation-based benchmarks are often unavailable at high dimensionality, ensuring the numerical stability and internal consistency of tomographic analytical models remains a key difficulty.

While this work focused on the cross-covariance between aperture mass variance and skewness, the combination $\Hat{\mathcal{M}}_\mathrm{ap}^2\times\Hat{\mathcal{M}}_\mathrm{ap}^3$ is not the optimal choice for joint second- and third-order statistics in real data analyses. A primary limitation is that aperture mass statistics require knowledge of the two- and three-point correlation functions down to arbitrarily small angular separations and, formally, up to infinite separation. In practice, this is problematic as close galaxy pairs may be blended or lost to detection limits in real data and finite resolution and pixelisation prevent accurate small-scale correlation measurements in simulations. These limitations can lead to E-/B-mode leakage and loss of accuracy. A more robust alternative is to use COSEBIs (Complete Orthogonal Sets of E-/B-mode Integrals) for the second-order component \citep{Schneider_2010}. COSEBIs are designed to operate over a finite angular range $[\theta_\mathrm{min},\theta_\mathrm{max}]$, with filter functions that are both complete and orthogonal. This ensures the mathematically exact separation of the E- and B-modes. While some residual leakage may occur due to numerical effects or masking, COSEBIs are considerably less sensitive to the limitations described above. Furthermore, COSEBIs offer a natural data compression: typically, the first 6 modes capture the majority of the cosmological information, even in tomographic analyses \citep{wright_2025}. A promising next step is therefore the development of an analytical cross-covariance between COSEBIs and $\Hat{\mathcal{M}}_\mathrm{ap}^3$, which would combine robustness, compactness, and sensitivity to non-Gaussian features. Additionally, second-order cosmic shear studies have also used the shear correlation functions $\xi_\pm$, so a joint $\xi_\pm \times \mathcal{M}_\mathrm{ap}^3$ covariance would also be a valuable extension.

\subsection{Summary}
\label{sec: disc summary}
In this work, we presented the first analytical real-space calculation of the cross-covariance between second- and third-order aperture mass statistics, completing the joint covariance matrix required for cosmological parameter estimation. For upcoming Stage IV surveys, analytical covariance matrices offer several advantages over simulation-based ones, including flexibility, scalability, and reduced noise. These benefits are most effectively realised when combined with COSEBIs for second-order statistics, and when extended to tomographic configurations. Nonetheless, limitations including the approximations inherent in the halo model, numerical instabilities in matrix operations, and difficulties in validation continue to hinder the broader adoption of analytical covariance matrices. Further developments are required to improve model accuracy, reduce numerical sensitivity, and enabling a tomographic set-up.

\begin{acknowledgements}
We would like to thank Joachim Harnois-D\'eraps for making public the SLICS mock data, which can be found at \url{http://slics.roe.ac.uk/}. We thank the anonymous referee for the useful comments. Funded by the TRA Matter (University of Bonn) as part of the Excellence Strategy of the federal and state governments. This work has been supported by the Deutsche Forschungsgemeinschaft through the project SCHN 342/15-1 and DFG SCHN 342/13. LL is supported by the Austrian Science Fund (FWF) [ESP 357-N]. PAB and SH acknowledge support from the German Academic Scholarship Foundation. LP acknowledges support from the DLR grant 50QE2002. SH is supported by the US Department of Energy, Office of High Energy Physics under Award Number DE-SC0019301.
\end{acknowledgements}

\bibliography{bibliography.bib}

\begin{appendix}

\section{The tetraspectrum halo model}
\label{app: tetra}

In this appendix, we approximate the tetraspectrum using a halo model as in \cite{Cooray2002}. The purpose of this estimation is to model $T_\mathrm{P5}$ (Eq.~\ref{eq: T4}), which can be slightly modified and expressed as
\begin{align}
    T_\mathrm{P5}(\theta_1,\theta_2,\theta_3;\theta_4)=&\;\int \frac{\dd^2\ell_1}{(2\pi)^2}\int \frac{\dd^2\ell_2}{(2\pi)^2} \int \frac{\dd^2s}{(2\pi)^2} \int \frac{\dd^2\ell_4}{(2\pi)^2}  \; P_5(\pmb{\ell}_1,\pmb{\ell}_2,\pmb{s}-\pmb{\ell}_1 - \pmb{\ell}_2,\pmb{\ell}_4)  \, \tilde{u}(\ell_1 \theta_1) \, \tilde{u}(\ell_2 \theta_2) \nonumber\\ &\times \tilde{u}(\lvert \pmb{s}-\pmb{\ell}_1 - \pmb{\ell}_2\rvert \ \theta_3)  \, \tilde{u}(\ell_4 \theta_4) \, \tilde{u}(\lvert \pmb{s}+\pmb{\ell}_4\rvert \, \theta_4)  \, G_A(\pmb{s})\;,
    \label{eq: T_4 halos}
\end{align}
where $\pmb{s}=\pmb{\ell}_1+\pmb{\ell}_2+\pmb{\ell}_3$ was used, such that the geometry factor only depends on a single input value. The tetraspectrum of the convergence is given by the Limber integration (Eq.~\ref{eq: limber}),
\begin{equation}
    P_5(\pmb{\ell}_1,\pmb{\ell}_2,\pmb{s}-\pmb{\ell}_1 - \pmb{\ell}_2,\pmb{\ell}_4)\,  = \left(\frac{3H_0^2\Omega_\mathrm{m}}{2c^2}\right)^5\,\int_0^\infty \dd{\chi}\; \frac{q^5(\chi)}{\chi^{3}\, a^5(\chi)}\, \mathcal{P}_5^\mathrm{(3d)}(\pmb{\ell}_1/\chi, \pmb{\ell}_2/\chi,(\pmb{s}-\pmb{\ell}_1-\pmb{\ell}_2)/\chi,\pmb{\ell}_4/\chi, (-\pmb{s}-\pmb{\ell}_4)/\chi; \chi)\;,
\end{equation}
where the tetraspectrum of the density contrast can be decomposed into terms that include up to five halos. Specifically, the one-halo term corresponds to all five points in the correlation function lying within a single halo, while the two-halo term involves the points being distributed between two halos. As a result, the tetraspectrum of the density contrast can be expressed as
\begin{align}
    \mathcal{P}_5^\mathrm{(3d)}&(\pmb{\ell}_1/\chi, \pmb{\ell}_2/\chi,(\pmb{s}-\pmb{\ell}_1-\pmb{\ell}_2)/\chi,\pmb{\ell}_4/\chi, (-\pmb{s}-\pmb{\ell}_4)/\chi; \chi)=I^\mathrm{0}_\mathrm{5}(\pmb{\ell}_1/\chi, \pmb{\ell}_2/\chi,(\pmb{s}-\pmb{\ell}_1-\pmb{\ell}_2)/\chi,\pmb{\ell}_4/\chi, (-\pmb{s}-\pmb{\ell}_4)/\chi; \chi) \nonumber\\ 
    &+ P_\mathrm{L}^\mathrm{(3d)}(\ell_1/\chi)\, I^\mathrm{1}_\mathrm{1}(\pmb{\ell}_1/\chi; \chi)\, I^\mathrm{1}_\mathrm{4}(\pmb{\ell}_2/\chi,(\pmb{s}-\pmb{\ell}_1-\pmb{\ell}_2)/\chi,\pmb{\ell}_4/\chi, (-\pmb{s}-\pmb{\ell}_4)/\chi; \chi)+ \mathrm{4\,Perm.} \nonumber\\ &+ P_\mathrm{L}^\mathrm{(3d)}(s/\chi)\, I^\mathrm{1}_\mathrm{3}(\pmb{\ell}_1/\chi, \pmb{\ell}_2/\chi, (\pmb{s}-\pmb{\ell}_1-\pmb{\ell}_2)/\chi;\chi)\, I^\mathrm{1}_\mathrm{2}(\pmb{\ell}_4/\chi,(\pmb{s}-\pmb{\ell}_4)/\chi; \chi)+ \mathrm{9\,Perm.} \label{eq: halos} \\ &+ \mathrm{3\,halo\,terms} + \mathrm{4\,halo\,terms} + \mathrm{5\,halo\,terms}\; , \nonumber 
\end{align}
where $P_\mathrm{L}^\mathrm{(3d)}$ is the linear density contrast power spectrum, and the $I_n^l$ terms are defined as
\begin{equation}
    I^l_n(\pmb{k}_1, ..., \pmb{k}_n; \chi)=\int \dd m\; n\left[m,z(\chi)\right]\, \left(\frac{m}{\Bar{\rho}}\right)^n\, b^l\left[m,z(\chi)\right]\, \prod_{i=1}^n\, \tilde{u}_\mathrm{NFW}(\pmb{k}_i,m)\; ,
    \label{eq: I terms}
\end{equation}
where $m$ is the halo mass, $n(m,z)$ is the halo mass function from \cite{Sheth1999}, $\Bar{\rho}$ is the comoving mean density, $b(m,z)$ the halo bias \citep{Sheth1999}, and $\tilde{u}_\mathrm{NFW}$ is the Fourier transformed and normalised Navarro–Frenk–White-halo profile, which is defined as $u_\mathrm{NFW}(r,m)=\rho(r,m)/m$, where $\rho$ is the halo density profile. We note that when $I_n^l$ terms are multiplied (meaning multiple halos), the masses, and thus also the integrations, are distinct per halo. 

As the distance between halos is much greater than within one halo, the one-halo term will dominate on small scales (large $\pmb{\ell}/\chi$). However, in the one-halo term, we only have correlations up to the scale of the largest halo. Therefore, we expect the one-halo term to be subdominant on scales larger than halos. It is shown in \cite{Pyne2021}, that on large scales, in the assumption that $s\ll \ell_1,\ell_2,\ell_4$, the two-halo term depending on $P_\mathrm{L}^\mathrm{(3d)}(s/\chi)$ dominates. Therefore, we can approximate Eq.~\eqref{eq: halos} as
\begin{align}
    \mathcal{P}_5^\mathrm{(3d)}&(\pmb{\ell}_1/\chi, \pmb{\ell}_2/\chi,(\pmb{s}-\pmb{\ell}_1-\pmb{\ell}_2)/\chi,\pmb{\ell}_4/\chi, (-\pmb{s}-\pmb{\ell}_4)/\chi; \chi)\approx I^\mathrm{0}_\mathrm{5}(\pmb{\ell}_1/\chi, \pmb{\ell}_2/\chi,(\pmb{s}-\pmb{\ell}_1-\pmb{\ell}_2)/\chi,\pmb{\ell}_4/\chi, (-\pmb{s}-\pmb{\ell}_4)/\chi; \chi) \nonumber\\ 
     &+ P_\mathrm{L}^\mathrm{(3d)}(s/\chi)\, I^\mathrm{1}_\mathrm{3}(\pmb{\ell}_1/\chi, \pmb{\ell}_2/\chi, (\pmb{s}-\pmb{\ell}_1-\pmb{\ell}_2)/\chi;\chi)\, I^\mathrm{1}_\mathrm{2}(\pmb{\ell}_4/\chi,(\pmb{s}-\pmb{\ell}_4)/\chi; \chi) \nonumber \\
     \approx&\; I^\mathrm{0}_\mathrm{5}(\pmb{\ell}_1/\chi, \pmb{\ell}_2/\chi,(-\pmb{\ell}_1-\pmb{\ell}_2)/\chi,\pmb{\ell}_4/\chi, -\pmb{\ell}_4/\chi; \chi) + P_\mathrm{L}^\mathrm{(3d)}(s/\chi)\, I^\mathrm{1}_\mathrm{3}(\pmb{\ell}_1/\chi, \pmb{\ell}_2/\chi, (-\pmb{\ell}_1-\pmb{\ell}_2)/\chi;\chi)\, I^\mathrm{1}_\mathrm{2}(\pmb{\ell}_4/\chi,-\pmb{\ell}_4/\chi; \chi)\; ,
\end{align}
where in the second approximation, we neglected $\pmb{s}$ except for in the linear power spectrum, as $P_\mathrm{L}^\mathrm{(3d)}(0)=0$. Now, filling in this approximation into Eq.~\eqref{eq: T_4 halos}, and using isotropy, yields
\begin{align}
    T_\mathrm{P5}(\theta_1,\theta_2,\theta_3;\theta_4)\approx &\; \left(\frac{3H_0^2\Omega_\mathrm{m}}{2c^2}\right)^5\,\int \frac{\dd^2\ell_1}{(2\pi)^2}\int \frac{\dd^2\ell_2}{(2\pi)^2} \int \frac{\dd^2s}{(2\pi)^2} \int \frac{\dd^2\ell_4}{(2\pi)^2}  \,\int_0^\infty \dd{\chi}\; \frac{q^5(\chi)}{\chi^{3}\, a(\chi)^5} \nonumber\\ &\times \tilde{u}(\ell_1 \theta_1) \, \tilde{u}(\ell_2 \theta_2) \, \tilde{u}(\lvert \pmb{\ell}_1 + \pmb{\ell}_2\rvert \ \theta_3)  \, \tilde{u}(\ell_4 \theta_4) \, \tilde{u}(\ell_4 \, \theta_4)  \, G_A(\pmb{s}) \nonumber \\ &\times \left[I^\mathrm{0}_\mathrm{5}(\ell_1/\chi, \ell_2/\chi,\lvert \ell_1+\ell_2\rvert/\chi,\ell_4/\chi, \ell_4/\chi; \chi) + P_\mathrm{L}^\mathrm{(3d)}(s/\chi)\, I^\mathrm{1}_\mathrm{3}(\ell_1/\chi, \ell_2/\chi, \lvert\ell_1+\ell_2\rvert/\chi;\chi)\, I^\mathrm{1}_\mathrm{2}(\ell_4/\chi,\ell_4/\chi; \chi)\right] \\ =&\; T_\mathrm{P5,1h}(\theta_1,\theta_2,\theta_3;\theta_4)+T_\mathrm{P5,2h}(\theta_1,\theta_2,\theta_3;\theta_4)\; , \nonumber
\end{align}
where
\begin{multline}
    T_\mathrm{P5,2h}(\theta_1,\theta_2,\theta_3;\theta_4)=\left(\frac{3H_0^2\Omega_\mathrm{m}}{2c^2}\right)^5\,\int_0^\infty \dd{\chi}\; \frac{q^5(\chi)}{\chi^{3}\, a(\chi)^5}
    \left[\int \frac{\dd^2s}{(2\pi)^2}\,G_A(\pmb{s}) \, P_\mathrm{L}^\mathrm{(3d)}(s/\chi)\right]\,\int \frac{\dd^2\ell_1}{(2\pi)^2}\int \frac{\dd^2\ell_2}{(2\pi)^2} \int \frac{\dd^2\ell_4}{(2\pi)^2}  \; \\ \times
     \tilde{u}(\ell_1 \theta_1) \, \tilde{u}(\ell_2 \theta_2) \, \tilde{u}(\lvert \pmb{\ell}_1 + \pmb{\ell}_2\rvert \ \theta_3)  \, \tilde{u}(\ell_4 \theta_4) \, \tilde{u}(\ell_4 \, \theta_4) \, I^\mathrm{1}_\mathrm{3}(\ell_1/\chi, \ell_2/\chi, \lvert\ell_1+\ell_2\rvert/\chi;\chi)\, I^\mathrm{1}_\mathrm{2}(\ell_4/\chi,\ell_4/\chi; \chi)
    \label{eq: T_4 with 2halo}
\end{multline}
will be zero in the large-field limit, where the integration over the Dirac delta function in Eq.~\eqref{eq: large field approx} gives us a factor $P_\mathrm{L}^\mathrm{(3d)}(0)=0$. Thus, we find a finite-field term similar to $T_\mathrm{PB,2}$.

\end{appendix}

\end{document}